\documentclass[12pt]{article}
\usepackage{axodraw}
  \newcommand{\ccaption}[2]{
    \begin{center}
    \parbox{0.85\textwidth}{
      \caption[#1]{\small{#2}}
      }
    \end{center}
    }

\def\Eqn#1{Eq.~(\ref{#1})}

\def\Eqns#1#2{Eqs.~(\ref{#1}) and (\ref{#2})}

\newcommand{\be}{\begin{equation}}
\newcommand{\ee}{\end{equation}}
\newcommand{\bea}{\begin{eqnarray}}
\newcommand{\eea}{\end{eqnarray}}
\newcommand{\OO}{{\cal O}}
\newcommand{\J}{{\cal J}}

\newcommand{\M}{{\cal M}}

\newcommand{\sss}[1]{\mbox{\scriptsize #1}}
\newcommand{\LL}{{\cal L}}
\newcommand{\PP}{{\cal P}}
\newcommand{\DD}{{\cal D}}
\newcommand{\Li}{\mbox{${\mbox L}{\mbox i}_{2}$}}
\newcommand{\peven}{$(V^2+A^2)$}
\newcommand{\podd}{$(VA)$}
\newcommand{\gev}{\mbox{GeV}}

\newcommand{\Tr}{\mbox{Tr}\,}

\newcommand{\cov}[2]{\mbox{\bf #1}_{\mbox{\tiny{\bf #2}}}}

\newcommand{\Lo}{\Lambda_1}
\newcommand{\Lt}{\Lambda_2}

\begin{document}

\begin{titlepage}

\begin{flushright}
  IPPP/01/35 \\
  DCPT/01/70 \\
  August 2001 \\ 
\end{flushright}
\vspace*{5mm}
\begin{center}
  {\bf NON-FACTORIZABLE CORRECTIONS AND EFFECTIVE FIELD THEORIES} \\
   \vspace*{2cm} 
{\bf A.P.~Chapovsky}, \ \  
{\bf V.A.~Khoze}, \ \  
{\bf A.~Signer} \ \  
{\bf and} \ \  
{\bf W.J. Stirling}\\ 
\vspace{0.6cm}
{\it
Institute for Particle Physics Phenomenology \\ 
University of Durham \\
Durham, DH1 3LE, England \\ }
  \vspace*{3cm}  
  {\bf ABSTRACT} \\ 
\end{center}
\vspace*{5mm}
\noindent
We analyze the structure of higher-order radiative corrections for
processes with unstable particles.  By subsequently integrating out
the various scales that are induced by the presence of unstable
particles we obtain a hierarchy of effective field theories.  In the
effective field theory framework the separation of physically
different effects is achieved naturally. In particular, we
automatically obtain a separation of factorizable and non-factorizable
corrections to all orders in perturbation theory. At one loop this
treatment is equivalent to the double-pole approximation (DPA) but
generalizes to higher orders and, at least in principle, to beyond the
DPA.  It is known that one-loop non-factorizable corrections to
invariant mass distributions are suppressed at high energy.  We study
the mechanism of this suppression and obtain estimates of higher-order
non-factorizable corrections at high energy.

\end{titlepage}


\section{Introduction}

An important area in the physics programme of a future
electron-positron linear collider (LC) belongs to the detailed studies
of the pair production of heavy unstable particles ($W$ and $Z$
bosons, top quarks, SUSY particles, etc.) \cite{zerwas}.  One of the
main goals of these measurements is the high precision determination
of their parameters, primarily their masses.

In recent years there has been a growing interest in high energy
photon colliders (see e.g.~\cite{telnov}), which in some cases offer
certain advantages over the $e^+e^-$ collisions in the exploration of
the unstable particles, see for example \cite{cks-rad}.  Pairs of
unstable particles can also be produced in hadron collisions at the
Tevatron and LHC. All these studies would allow a variety of detailed
tests of the Standard Model to be performed, both in the electroweak
and in the strong interaction sectors. This, in turn, requires very
accurate theoretical knowledge of the production and decay mechanism
and of interference effects between them.  This interference is caused
by the large width, $\Gamma$, of many of these objects.  The widths of
the $W$, $Z$ and $t$ are all of the order of $2~\gev$, and many
supersymmetric and other beyond-the-Standard-Model particles would
also have widths in the $\gev$ range.

Although many results of this paper are applicable to the
non-relativistic case as well, we concentrate on relativistic and
ultra-relativistic energy regimes. In the center-of-mass frame the
physical picture of processes involving unstable particles can be
viewed as a consequence of several subprocesses. First the unstable
particles are produced during a short time scale $\OO(1/E)$. Then
during a large proper time, $\sim 1/\Gamma$, they separate from each
other on a large distance, $\sim 1/\Gamma \cdot E/M$. Finally, they
decay during a short time, $\sim 1/E$, into their decay
products. Here, $E,M$ and $\Gamma$ are the energy, mass and width of
the unstable particle respectively. If $E^2\gg M\Gamma$, the
production and decay subprocesses are well separated from each other.
The so-called pole-scheme~\cite{dpa} is a framework which makes use of
the presence of two scales in the problem by expanding the amplitudes
in $\Gamma/M$.  In particular, one can keep only the leading terms,
neglecting all the terms suppressed by powers of $\Gamma/M$.  In the
case of pair production of unstable particles such an approximation is
usually called the double-pole approximation (DPA).  The DPA
guarantees gauge invariance of the calculation and simplifies it
significantly.  The DPA allowed the calculation of the full
$\OO(\alpha)$ electroweak corrections to the pair production of $W$
and $Z$ bosons in $e^+e^-$ collisions~\cite{dpa-calc}.  It is
essential for the DPA that unstable particles are close to resonance
and that $E^2\gg M\Gamma$.  This is also the regime which we will
specialize to throughout this paper.

Within the DPA at one loop one can readily classify the radiative
effects into two types: {\it factorizable}, which act inside the
separate hard subprocesses (production and decay); and {\it
non-factorizable}, which interconnect various hard subprocesses.
Depending on the subprocesses they interconnect, the non-factorizable
corrections can be of two types: decay-decay and production-decay.
The non-factorizable corrections are believed to to be an important
ingredient of the theoretical predictions.  In particular, they affect
the measurement of the mass of the unstable particle.  They have been
a subject of intensive study recently, see for example \cite{cks-rad},
\cite{kos-rad}-\cite{sjos-nonpert-tt}.

The properties of perturbative non-factorizable corrections depend
crucially on the level of ``inclusiveness'' of the distributions.  The
distributions can be of three distinct types.  For virtual corrections
only two of them are relevant: invariant mass and angular
distributions. Since real gauge-boson (photon, gluon) radiation
contributes to the non-factorizable corrections as well one can
consider additional distributions where the non-factorizable real
radiation is added and an integration over the real emission phase
space is performed.  At one loop in QED the non-factorizable
corrections to the distributions inclusive in real photon and
invariant mass are suppressed by at least $\alpha\Gamma/M$,
\cite{nf-theorem/fadin, nf-theorem/my}.  In
\cite{nf-theorem/fadin} general arguments are given indicating that
this should still be the case at higher loops.  As a consequence, for
a particular class of the one-loop non-factorizable corrections
(initial-final state interferences) the virtual contribution is
cancelled by the real emission even for the distributions exclusive in
the invariant mass.  In \cite{nf-calc/my} a first calculation of
one-loop non-factorizable corrections exclusive in invariant mass but
inclusive in real radiation was performed.  In \cite{nf-calc/bbc,
nf-calc/ddr} the one-loop non-factorizable corrections were calculated
for completely exclusive distributions.  Although the latter
calculations were originally performed for pair production of
$W$ bosons, the results were applied also to other cases, for example
to one-loop QCD non-factorizable corrections in top-quark pair
production~\cite{nf-tt} and to one-loop QED non-factorizable
corrections to the pair production of $Z$ bosons~\cite{nf-zz}.

Even though one-loop non-factorizable corrections are extensively
studied by now, almost nothing is known about the higher-loop
non-factorizable effects.  An explicit calculation of two-loop
non-factorizable corrections is well beyond present capabilities.
Even a precise definition of higher-loop non-factorizable corrections
is lacking at present. Higher-order non-factorizable corrections can
be quite important, especially in QCD.  Note that all of the one-loop
results above did not assume the distributions to be inclusive in the
angles.  In \cite{nf-anz} it was found that the one-loop
non-factorizable correction to the distribution inclusive in real
photons and angles, but not in the invariant mass, is suppressed by
$\alpha (M/E)^4$ at high energies, $E\gg M$. This is consistent with
an earlier observation of Ref.~\cite{nf-calc/my}, where a similar
effect was observed for a simplified toy model. A similar energy
dependence is also found for the one-loop QCD interconnection effects
in $t\bar{t}$ production~\cite{nf-tt,nf-anz}.  It is the main purpose
of this paper to study the mechanism of the high energy
suppression. This allows us to obtain information about the
higher-order non-factorizable corrections to the distributions
inclusive in angles.

An indication of the importance of higher-loop QCD corrections comes
from the studies of non-perturbative QCD interconnection effects.
These effects are an essential source of the systematic uncertainties
in the reconstruction of the $W$-boson or top-quark parameters.
Recall that not far from threshold the typical decay time,
$\tau\sim1/\Gamma\approx 0.1 \mbox{fm}$, is much shorter than the
characteristic hadronization time, $\tau{\sss{had}}\approx
1~\mbox{fm}$.  Thus, the final state hadronic systems overlap between
pairs of resonances ($W^+W^-$, $Z^0Z^0$, $t\bar{t}$, etc).
Non-perturbative effects at the hadronization stage are usually
modelled by a colour rearrangement between the partons produced in the
two resonance decays and the subsequent parton showers
\cite{sjos-nonpert}.  This topic is deeply related to the confinement
physics.  Thus, interconnection studies not only offer an opportunity
to investigate the dynamics of unstable particles, but they also open
new ways to probe confinement forces in space and time
\cite{sjos-nonpert,sjos-nonpert-x}.

So far, the experimental analysis of the non-factorizable effects have
been performed only at LEP2 and mainly in the context of precise $W$
mass measurements~\cite{nf-exp}.  A LC will allow a series of very
accurate measurements of $W^+W^-$, $ZZ$, and $t\bar{t}$ production in
a wide energy range. Especially promising for the purposes of precise
mass determination and interconnection studies look high luminosity
runs of a LC in the threshold regions \cite{sjos-nonpert-x,
sjos-nonpert-tt}. A detailed knowledge of the energy dependence of
the interconnection effects in $W^+W^-$ production would allow to
choose the optimal strategy for their studies at a LC.

The purpose of this paper is twofold.  First, we discuss the
separation between the different radiative phenomena to all orders in
perturbation theory.  We find the following different effects:
\begin{itemize}
	\item There are factorizable and non-factorizable corrections.
	\item The non-factorizable corrections can be of the
	production-decay and decay-decay types.  
        \item The non-factorizable corrections of each type can receive
	corrections due to two effects: interaction between the
	production/decay dipoles and propagation corrections.
\end{itemize}
We start with the separation of these effects in QED within the DPA by
analyzing Feynman diagrams. This is an extension of the well studied
one-loop QED case.  Subsequently we find an effective field theory
interpretation of this separation, based on the presence of a
hierarchy of scales in the problem, $E^2\gg M\Gamma$.  The existence
of an effective field theory allows us to generalize the separation
between different effects to the more complicated (non-abelian) QCD
case.  Also the DPA appears naturally within the effective field
theory framework.  This implies -- at least in principle -- the
possibility to study $\Gamma/M$ suppressed (beyond the DPA) effects in
a consistent way.

Second, we focus on the high-energy, $E\gg M$, behavior of the
non-factorizable corrections to the distributions inclusive in angles
of the decay products.  We study the mechanism responsible for the
suppression of the non-factorizable corrections at high energy.  The
suppression mechanism works differently for the different effects we
identified.  In this paper we limit ourselves to the decay-decay
interferences, leaving aside the more complicated case of the
production-decay interferences.  We illustrate the suppression
mechanism by estimating the QED decay-decay interferences, using a
high energy expansion of Feynman diagrams.  Subsequently we find an
effective field theory interpretation of this mechanism.  Based on the
effective field theory approach we develop a general framework that
allows us to generalize the estimates to the non-abelian QCD
case. Finally, we consider some topical applications of our results.

The paper is organized as follows.  In Sect.~\ref{sec:feyn} we discuss
the separation of different effects in QED by analysing Feynman
diagrams.  We also derive the high energy estimates of the
non-factorizable decay-decay interferences.  In Sect.~\ref{sec:eff} we
analyse the structure of the modes induced by the hierarchy of scales,
$E$ and $\Gamma$.  By subsequently integrating out different modes we
construct an effective field theory that provides us with a separation
between the various radiative effects together with the high energy
estimates.  In Sect.~\ref{sec:appl} we discuss some physical
applications of the high energy estimates of the non-factorizable
corrections.  Sect.~\ref{sec:concl} gives our conclusions and outlines
the possible future developments and applications of our results.


\section{Estimates from high energy expansion of amplitudes }
\label{sec:feyn}

In this section we investigate higher-loop QED non-factorizable
corrections in a toy model by considering the high-energy expansion
of Feynman diagrams.  We work within the DPA, neglecting all
contributions suppressed by $\Gamma/M$.  The separation of
factorizable and non-factorizable corrections is achieved by using the
non-factorizable currents.  This procedure is an extension of the
one-loop separation procedure~\cite{dpa-calc, nf-calc/bbc,
nf-calc/ddr}.  We subsequently study non-factorizable corrections to
the distributions inclusive in decay angles in the high energy limit,
$E\gg M$.

\subsection{Separation of various contributions}
\label{sec:factorization}

In the Born approximation the hierarchy of scales $E^2\gg M \Gamma$
implies the factorization of the full matrix element into a product of
matrix elements corresponding to hard (production and decay of
unstable particles) and soft (propagation of unstable particles)
sub-stages of the process. The factorization holds with an accuracy of
at least $\OO(\Gamma/M)$.

Radiative corrections are due to radiation of photons/gluons with
typical energy $\Omega$.  If the radiation is hard, $\Omega\sim E$,
the inverse propagation distance of the unstable particles, $\sim
\Gamma \cdot M/E$, remains to be the only soft scale in the
problem. As a result, the factorization for the process works in the
same way as in the Born approximation. On the other hand, if the
radiation is soft, $\Omega\sim \Gamma \cdot M/E$, there are two soft
scales. The hard matrix elements still factorize but the propagation
of the unstable particles and the interactions via soft particles can
mix now. In general, a part of the corrections will have a form where
the propagation subprocess is factorized in the same way as it is
factorized in the Born approximation, and a part of the corrections
will not have such a factorized form.

A correction is factorizable (non-factorizable) if the propagation
sub-stage factorizes (does not factorize)with respect to the rest of
the process in the same way as in the Born approximation. Corrections
due to the exchange of hard particles are always factorizable whereas
corrections due to the exchange of soft particles can be both
factorizable and non-factorizable. In the former case the split up
between factorizable and non-factorizable corrections is unambiguously
defined on the basis of the energy of the exchange particles. In the
latter case the split up between factorizable and non-factorizable
corrections can only be done by comparing corresponding matrix
elements to that in the Born approximation.  One expects that
factorizable and non-factorizable contributions factorize with respect
to each other beyond one loop.

The toy model we will consider has a neutral, scalar, unstable
particle field $\phi$ with mass $M$ and width $\Gamma$. This particle
decays into massless, charged fermions, $\psi$. The charged fermions
couple to a $U(1)$ gauge field, $A_\mu$.  The Lagrangian of the model
is
\be
\label{feyn/model}
	\LL =
	\frac{1}{2} \phi \bigl(p^2-M^2\bigr) \phi 
	+ \bar{\psi} \not\! \cov{p}{} \psi
        - \frac{1}{4} F_{\mu\nu} F^{\mu\nu}
	+  \PP X\phi\phi 
	+ \phi\bar{\psi} \DD \psi.
\ee
Here $p$ is the momentum operator and $\cov{p}{}$ is the covariant
momentum operator. The first three terms in \Eqn{feyn/model} describe
the propagation of $\phi$, $\psi$, and $A_\mu$ fields. The $\PP$ term
describes the production of a pair of unstable particles from some
neutral source $X$, whereas the $\DD$ term describes their decay into
the fermions. We will assume that all these terms can be treated as a
perturbation. To see how factorization occurs let us consider the
process
\be
  X \to \phi(p_1) \phi(p_2) \to 
  \psi^+(k_1)\psi^-(k_1') \psi^-(k_2)\psi^+(k_2').
  \label{toy-process}
\ee
where $p_i = k_i+k_i', \ i\in\{1,2\}$ and  
\be
p_1^2-M^2\sim \Gamma M, \quad  
p_2^2-M^2\sim \Gamma M
\label{kin-const}
\ee
We assume there are no charged particles in the initial state.

\subsubsection{One-loop order}

The complete one-loop correction consists of one-loop factorizable and
one-loop non-factorizable corrections. The separation between these
two contributions is known~\cite{dpa-calc}. In order to introduce
notations let us recall briefly how it works. The hard modes
contribute exclusively to the factorizable corrections, thus the
separation needs to be performed only for soft photons.  Let us start
with the non-factorizable corrections. Since charged particles appear
only in the final state, as decay products there will be only
decay-decay interference.  The decay-decay non-factorizable correction
is given by
\be
\label{feyn/1loop/nf}
 \begin{picture}(60,30)(0,-2) 
	\GCirc(0,0){3}{0}
	\DashLine(0,0)(30,20){3}
	\DashLine(0,0)(30,-20){3}
	\GCirc(30,20){3}{0}
	\ArrowLine(30,20)(60,30)
	\ArrowLine(60,10)(30,20)
	\GCirc(30,-20){3}{0}
	\ArrowLine(30,-20)(60,-10)
	\ArrowLine(60,-30)(30,-20)
	\Photon(30,20)(30,-20){2}{6}
\end{picture}
	\ = \ \ 
\begin{picture}(30,30)(0,-2) 
	\GCirc(0,0){3}{0}
	\DashLine(0,0)(0,20){3}
	\DashLine(0,0)(0,-20){3}
	\GCirc(0,20){2}{0}
	\ArrowLine(0,20)(30,30)
	\ArrowLine(30,10)(00,20)
	\GCirc(0,-20){2}{0}
	\ArrowLine(0,-20)(30,-10)
	\ArrowLine(30,-30)(0,-20)
	\Photon(15,15)(15,-15){2}{6}
\end{picture}
	\ + \ \ 
\begin{picture}(30,30)(0,-2) 
	\GCirc(0,0){3}{0}
	\DashLine(0,0)(0,20){3}
	\DashLine(0,0)(0,-20){3}
	\GCirc(0,20){2}{0}
	\ArrowLine(0,20)(30,30)
	\ArrowLine(30,10)(00,20)
	\GCirc(0,-20){2}{0}
	\ArrowLine(0,-20)(30,-10)
	\ArrowLine(30,-30)(0,-20)
	\Photon(15,25)(15,-15){2}{7}
\end{picture}
 	\ + \  \ 
\begin{picture}(30,30)(0,-2) 
	\GCirc(0,0){3}{0}
	\DashLine(0,0)(0,20){3}
	\DashLine(0,0)(0,-20){3}
	\GCirc(0,20){2}{0}
	\ArrowLine(0,20)(30,30)
	\ArrowLine(30,10)(00,20)
	\GCirc(0,-20){2}{0}
	\ArrowLine(0,-20)(30,-10)
	\ArrowLine(30,-30)(0,-20)
	\Photon(15,15)(15,-25){2}{7}
\end{picture}
 	\ + \ \ 
\begin{picture}(30,30)(0,-2) 
	\GCirc(0,0){3}{0}
	\DashLine(0,0)(0,20){3}
	\DashLine(0,0)(0,-20){3}
	\GCirc(0,20){2}{0}
	\ArrowLine(0,20)(30,30)
	\ArrowLine(30,10)(00,20)
	\GCirc(0,-20){2}{0}
	\ArrowLine(0,-20)(30,-10)
	\ArrowLine(30,-30)(0,-20)
	\Photon(15,25)(15,-25){2}{9}
\end{picture}
\rule[-30pt]{0mm}{0mm}
	\ = \ 
	i\M_{0}
	\int_{l}
	\ 
	\bigl(\J_{1} \cdot \J_{2}\bigr),
\ee
where we used a short hand notation for the loop integral
\be
	\int_{l} \equiv \int\frac{d^{4}l}{(2\pi)^{4} l^{2}}\ .
\label{intmeasure}
\ee
In \Eqn{feyn/1loop/nf} $\M_{0}$ is the Born matrix element in
the DPA.  The gauge invariant currents $\J_{1,2}^{\mu}$ are called the
non-factorizable currents.  They are given by
\bea
	\J_{1}^{\mu} &=& \J_{1}^{\mu}(l, D_1)
	= +\,e\Biggl[
                 \frac{k_{1}^{\mu}}{lk_{1}}
               - \frac{{k_{1}'}^{\mu}}{lk_{1}'}
        \Biggr]\frac{D_{1}}{D_{1}+2lp_{1}},
\label{feyn/currents} \\
	\J_{2}^{\mu} &=& \J_{2}^{\mu}(l, D_2)
	= +\,e\Biggl[
                  \frac{k_{2}^{\mu}}{-lk_{2}}
                - \frac{{k_{2}'}^{\mu}}{-lk_{2}'}
        \Biggr]\frac{D_{2}}{D_{2}-2lp_{2}},
\nonumber
\eea
where 
\be
	D_{1,2} \equiv p_{1,2}^{2}-M^2 + i \Gamma M,
\ee 
is the off-shellness of the unstable particle.  By inspection of the
integral \Eqn{feyn/1loop/nf} it can be seen that the energy of the
photon relevant for non-factorizable corrections is indeed
$\Omega\sim\Gamma M/E$. The contribution of a hard photon is
suppressed by $\OO(M^2 \Gamma^2/E^2)$ with respect to the leading
contribution.  Thus, the corresponding contributions are beyond the
DPA.

The factorizable correction is given by
\be
\begin{picture}(30,30)(0,-2) 
	\GCirc(0,0){3}{0}
	\DashLine(0,0)(0,20){3}
	\DashLine(0,0)(0,-20){3}
	\GCirc(0,20){2}{0}
	\ArrowLine(0,20)(30,30)
	\ArrowLine(30,10)(00,20)
	\GCirc(0,-20){2}{0}
	\ArrowLine(0,-20)(30,-10)
	\ArrowLine(30,-30)(0,-20)
	\Photon(20,26.666)(20,13.333){2}{3}
\end{picture}
 \ \ \ + \ \ \ 
\begin{picture}(30,30)(0,-2) 
	\GCirc(0,0){3}{0}
	\DashLine(0,0)(0,20){3}
	\DashLine(0,0)(0,-20){3}
	\GCirc(0,20){2}{0}
	\ArrowLine(0,20)(30,30)
	\ArrowLine(30,10)(00,20)
	\GCirc(0,-20){2}{0}
	\ArrowLine(0,-20)(30,-10)
	\ArrowLine(30,-30)(0,-20)
	\Photon(20,-26.666)(20,-13.333){2}{3}
\end{picture}
\rule[-30pt]{0mm}{0mm}
  \ \  \ + \ \ \ \ldots
	\ = \ 
	\M_1.
\ee
The important property of $\M_1$ is that it has exactly the same
dependence on $D_{1,2}$ as the Born matrix element $\M_0$.  This is
the crucial difference between the factorizable and non-factorizable
corrections.  The non-factorizable corrections contain an additional
strong $D_{1,2}$ dependence. Thus, contrary to the factorizable
corrections, the non-factorizable corrections distort the invariant
mass distribution.

\subsubsection{Two-loop order}

At two-loop the separation into factorizable and non-factorizable
contributions is somewhat more involved. There are four distinct
contributions, two of which are non-factorizable. First of all there
is a two-loop factorizable correction given by the following diagrams
\be
\begin{picture}(60,30)(0,-2) 
	\GCirc(0,0){3}{0}
	\DashLine(0,0)(30,20){3}
	\DashLine(0,0)(30,-20){3}
	\GCirc(30,20){2}{0}
	\ArrowLine(30,20)(60,30)
	\ArrowLine(60,10)(30,20)
	\GCirc(30,-20){2}{0}
	\ArrowLine(30,-20)(60,-10)
	\ArrowLine(60,-30)(30,-20)
	\Photon(50,26.666)(50,13.333){2}{3}
	\Photon(55,28.336)(55,11.6675){2}{3}
\end{picture}
 \ \ \ + \ \ \ 
\begin{picture}(60,30)(0,-2) 
	\GCirc(0,0){3}{0}
	\DashLine(0,0)(30,20){3}
	\DashLine(0,0)(30,-20){3}
	\GCirc(30,20){2}{0}
	\ArrowLine(30,20)(60,30)
	\ArrowLine(60,10)(30,20)
	\GCirc(30,-20){2}{0}
	\ArrowLine(30,-20)(60,-10)
	\ArrowLine(60,-30)(30,-20)
	\Photon(50,-26.666)(50,-13.333){2}{3}
	\Photon(55,-28.336)(55,-11.6675){2}{3}
\end{picture}
\rule[-30pt]{0mm}{0mm}
  \ \  \ + \ \ \ \ldots
	\ = \ 
	\M_2.
\ee

Secondly, there are contributions that can be attributed to
the interference between one-loop factorizable and one-loop
non-factorizable contributions.  An example of such a contribution in
the soft approximation is
\be 
\begin{picture}(30,30)(0,-2) 
	\GCirc(0,0){3}{0}
	\DashLine(0,0)(0,20){3}
	\DashLine(0,0)(0,-20){3}
	\GCirc(0,20){2}{0}
	\ArrowLine(0,20)(30,30)
	\ArrowLine(30,10)(00,20)
	\GCirc(0,-20){2}{0}
	\ArrowLine(0,-20)(30,-10)
	\ArrowLine(30,-30)(0,-20)
	\Photon(20,26.666)(20,13.333){2}{3}
	\Photon(10,16.666)(10,-16.666){2}{7}
\end{picture}
 \ \ \ + \ \ \ 
\begin{picture}(30,30)(0,-2) 
	\GCirc(0,0){3}{0}
	\DashLine(0,0)(0,20){3}
	\DashLine(0,0)(0,-20){3}
	\GCirc(0,20){2}{0}
	\ArrowLine(0,20)(30,30)
	\ArrowLine(30,10)(00,20)
	\GCirc(0,-20){2}{0}
	\ArrowLine(0,-20)(30,-10)
	\ArrowLine(30,-30)(0,-20)
	\Photon(10,+23.333)(10,16.666){2}{2}
	\Photon(20,13.333)(20,-13.333){2}{5}
\end{picture}
\rule[-30pt]{0mm}{0mm}
       \  = \ i\M_{0} \ 
	\int_{l_1, l_2} \ 
	\frac{k_1^{\mu}}{l_2k_1} \ \frac{D_1}{D_1+2l_2p_1}
	\ \cdot \ 
	\frac{k_{2 \mu}}{-l_2k_2} \ \frac{D_2}{D_2-2l_2p_2}
	\ \times \ 
	\frac{k_1^{\nu}}{l_1k_1}
	\ \cdot \ 
	\frac{k_{1 \nu}^{\prime}}{-l_1k_1'}.	
\ee
The two-loop corrections of this type factorize into a product of
one-loop factorizable corrections times one-loop non-factorizable
corrections. The same factorization holds for hard momenta flow in
the factorizable loop. Combination of all contributions of this type
will lead to
\be
 \begin{picture}(60,30)(0,-2) 
	\GCirc(0,0){3}{0}
	\DashLine(0,0)(30,20){3}
	\DashLine(0,0)(30,-20){3}
	\GCirc(30,20){3}{0}
	\ArrowLine(30,20)(60,30)
	\ArrowLine(60,10)(30,20)
	\GCirc(30,-20){3}{0}
	\ArrowLine(30,-20)(60,-10)
	\ArrowLine(60,-30)(30,-20)
	\Photon(30,20)(30,-20){2}{6}
	\Photon(50,26.666)(50,13.333){2}{3}
\end{picture} 
	\ + \ \ 
\begin{picture}(60,30)(0,-2) 
	\GCirc(0,0){3}{0}
	\DashLine(0,0)(30,20){3}
	\DashLine(0,0)(30,-20){3}
	\GCirc(30,20){3}{0}
	\ArrowLine(30,20)(60,30)
	\ArrowLine(60,10)(30,20)
	\GCirc(30,-20){3}{0}
	\ArrowLine(30,-20)(60,-10)
	\ArrowLine(60,-30)(30,-20)
	\Photon(30,20)(30,-20){2}{6}	
	\Photon(50,-26.666)(50,-13.333){2}{3}
\end{picture}
\rule[-30pt]{0mm}{0mm}
	\ = \ 
	i\M_{1} \ 
	\int_{l}
	\ 
	\bigl(\J_{1} \cdot \J_{2}\bigr).
\ee
The first factor, $\M_{1}$, is the same as the one-loop factorizable
correction.  The second factor is the same as one-loop
non-factorizable correction.

In addition to the two contributions mentioned above, there are also
two classes of non-factorizable contributions. The first class
consists of all diagrams where two photons are exchanged between the
two fermion pairs. Let us list here all these diagrams. There are
diagrams like
\be
\begin{picture}(30,30)(0,-2) 
	\GCirc(0,0){3}{0}
	\DashLine(0,0)(0,20){3}
	\DashLine(0,0)(0,-20){3}
	\GCirc(0,20){2}{0}
	\ArrowLine(0,20)(30,30)
	\ArrowLine(30,10)(00,20)
	\GCirc(0,-20){2}{0}
	\ArrowLine(0,-20)(30,-10)
	\ArrowLine(30,-30)(0,-20)
	\Photon(10,16.666)(10,-16.666){2}{7}
	\Photon(20,13.333)(20,-13.333){2}{5}
\end{picture}
	\ + \ \ 
\begin{picture}(30,30)(0,-2) 
	\GCirc(0,0){3}{0}
	\DashLine(0,0)(0,20){3}
	\DashLine(0,0)(0,-20){3}
	\GCirc(0,20){2}{0}
	\ArrowLine(0,20)(30,30)
	\ArrowLine(30,10)(00,20)
	\GCirc(0,-20){2}{0}
	\ArrowLine(0,-20)(30,-10)
	\ArrowLine(30,-30)(0,-20)
	\Photon(10,16.666)(20,-13.333){2}{7}
	\Photon(20,13.333)(10,-16.666){2}{7}
\end{picture} 
\rule[-30pt]{0mm}{0mm}
   \ \sim \ \ 
	\frac{1}{2} \ 
	\int_{l_1, l_2} \ 
	\frac{k_1^{\mu}}{k_1 l_1} \ \ 
	\frac{k_1^{\nu}}{k_1 l_2}	\ \ 
	\frac{k_{2 \mu}}{- k_2 l_1} \ \ 
	\frac{k_{2 \nu}}{- k_2 l_2},
\ee
where the integration is over the virtual photon momenta, $l_1$ and
$l_2$, with the appropriate measure, \Eqn{intmeasure}.  We omit
factors $D_{1,2}/(D_{1,2}\pm 2l_1p_{1,2}\pm 2l_2p_{1,2})$ in the
expression above.  There are three more contributions of this form
corresponding to interactions: $(k_1k_2')(k_1k_2')$,
$(k_1'k_2)(k_1'k_2)$, $(k_1'k_2')(k_1'k_2')$.  Another set of diagrams
contributing to this class is given by
\be
\begin{picture}(30,30)(0,-2) 
	\GCirc(0,0){3}{0}
	\DashLine(0,0)(0,20){3}
	\DashLine(0,0)(0,-20){3}
	\GCirc(0,20){2}{0}
	\ArrowLine(0,20)(30,30)
	\ArrowLine(30,10)(00,20)
	\GCirc(0,-20){2}{0}
	\ArrowLine(0,-20)(30,-10)
	\ArrowLine(30,-30)(0,-20)
	\Photon(10,23.333)(10,-16.666){2}{8}
	\Photon(20,13.333)(20,-13.333){2}{5}
\end{picture}
	\ + \ \ 
\begin{picture}(30,30)(0,-2) 
	\GCirc(0,0){3}{0}
	\DashLine(0,0)(0,20){3}
	\DashLine(0,0)(0,-20){3}
	\GCirc(0,20){2}{0}
	\ArrowLine(0,20)(30,30)
	\ArrowLine(30,10)(00,20)
	\GCirc(0,-20){2}{0}
	\ArrowLine(0,-20)(30,-10)
	\ArrowLine(30,-30)(0,-20)
	\Photon(10,16.666)(10,-16.666){2}{7}
	\Photon(20,-13.333)(20,+26.666){2}{9}
\end{picture} 
\rule[-30pt]{0mm}{0mm}
   \ \sim \ \ 
        -
	\frac{1}{2} \ 
	\int_{l_1, l_2} \ 
	\Biggl[
	\frac{k_1^{\mu}}{k_1 l_1} \ 
	\frac{k_1^{\prime \nu}}{k_1' l_2} \ 
	\frac{k_{2 \mu}}{- k_2 l_1} \ 
	\frac{k_{2 \nu}}{- k_2 l_2} \ 
	 \ \ + \ \ 
	\frac{k_1^{\prime \mu}}{k_1' l_1} \ 
	\frac{k_1^{\nu}}{k_1 l_2} \ 
	\frac{k_{2 \mu}}{- k_2 l_1}  \ 
	\frac{k_{2 \nu}}{- k_2 l_2}  \ 
	\Biggr].
\ee
There are three additional contributions of this type:
$(k_1k_2')(k_1'k_2')$, $(k_1k_2)(k_1k_2')$,
$(k_1'k_2)(k_1'k_2')$. Finally, the last set of diagrams contributing
to the first class of non-factorizable corrections is
\be
\begin{picture}(30,30)(0,-2) 
	\GCirc(0,0){3}{0}
	\DashLine(0,0)(0,20){3}
	\DashLine(0,0)(0,-20){3}
	\GCirc(0,20){2}{0}
	\ArrowLine(0,20)(30,30)
	\ArrowLine(30,10)(00,20)
	\GCirc(0,-20){2}{0}
	\ArrowLine(0,-20)(30,-10)
	\ArrowLine(30,-30)(0,-20)
	\Photon(10,23.333)(10,-16.666){2}{8}
	\Photon(20,13.333)(20,-26.666){2}{8}
\end{picture}
\rule[-30pt]{0mm}{0mm}
   \ \sim \ \ 
        + \ 
	\int_{l_1, l_2} \ 
	\frac{k_1^{\prime \mu}}{k_1' l_1} \ 
	\frac{k_1^{\nu}}{k_1 l_2} \ 
	\frac{k_{2 \mu}}{- k_2 l_1} \ 
	\frac{k_{2 \nu}^{\prime}}{- k_2' l_2}.
\ee
There is one more diagram of this type: $(k_1k_2)(k_1'k_2')$.

Combining all these contributions we obtain a gauge invariant
expression for this part of the two-loop non-factorizable correction
\bea
	\frac{1}{2} \ 
	i\M_{0} \! && 
	\int_{l_1,l_2} \ 
	\Biggl(\frac{k_1}{l_1k_1}-\frac{k_1'}{l_1k_1'}\Biggr)^{\mu}\cdot
	\Biggl(\frac{k_1}{l_2k_1}-\frac{k_1'}{l_2k_1'}\Biggr)^{\nu}   \cdot
	\Biggl(\frac{k_2}{-l_1k_2}-\frac{k_2'}{-l_1k_2'}\Biggr)_{\mu} \cdot
	\Biggl(\frac{k_2}{-l_2k_2}-\frac{k_2'}{-l_2k_2'}\Biggr)_{\nu} 
\nonumber \\
&& \qquad \qquad \qquad \qquad
	\frac{D_1}{D_1+2l_1p_1+2l_2p_1} \ \ 
	\frac{D_2}{D_2-2l_1p_2-2l_2p_2}
\eea
Using the non-factorizable currents this can be written as
\be
\label{feyn/2loop/nf}
 \begin{picture}(60,30)(0,-2) 
	\GCirc(0,0){3}{0}
	\DashLine(0,0)(30,20){3}
	\DashLine(0,0)(30,-20){3}
	\GCirc(30,20){3}{0}
	\ArrowLine(30,20)(60,30)
	\ArrowLine(60,10)(30,20)
	\GCirc(30,-20){3}{0}
	\ArrowLine(30,-20)(60,-10)
	\ArrowLine(60,-30)(30,-20)
	\PhotonArc(60,0)(36.05,150,210){2}{6}	
	\PhotonArc(0,0)(36.05,-30,30){2}{6}
\end{picture} 
\rule[-30pt]{0mm}{0mm} 
  \ \sim \ \ 
	\frac{1}{2} \ 
	i\M_{0} \ 
	\int_{l_1, l_2} \ 
	\bigl(\J_{1}\cdot\J_{2}\bigr)(l_1)
	*
	\bigl(\J_{1}\cdot\J_{2}\bigr)(l_2),
\ee
where the currents are the same as in the one-loop case and are given
by \Eqn{feyn/currents}.  The product of two currents is defined by
the following current multiplication rules
\bea
	\phantom{ = }
	\J_i^{\mu}(l_1,D) * \J_i^{\nu}(l_2,D)
	&=&
        \J_i^{\mu}(l_1,D) \ \J_i^{\nu}(l_2,D\pm2l_1p_i)	 
\nonumber \\
\label{feyn/prod_of_currents} 
	= 	
	\J_i^{\nu}(l_1,D) * \J_i^{\mu}(l_2,D)
	&=&
        \J_i^{\nu}(l_1,D) \ \J_i^{\mu}(l_2,D\pm2l_1p_i),
\eea
where $\pm=+$ for $i=1$ and $\pm=-$ for $i=2$. Furthermore
\be
\label{feyn/prod_of_currents_x}
	\J_1^{\mu}(l_1,D) * \J_2^{\nu}(l_2,D)
	=
	\J_2^{\nu}(l_2,D) * \J_1^{\mu}(l_1,D)
	=
	\J_1^{\mu}(l_1,D) \ \J_2^{\nu}(l_2,D).
\ee

The second class of non-factorizable corrections consists of diagrams
with only one photon exchange. These are diagrams of the following form
\be
\label{feyn/2loop/nf_x} 
\begin{picture}(60,30)(0,-2) 
	\GCirc(0,0){3}{0}
	\DashLine(0,0)(30,20){3}
	\DashLine(0,0)(30,-20){3}
	\GCirc(30,20){3}{0}
	\ArrowLine(30,20)(60,30)
	\ArrowLine(60,10)(30,20)
	\GCirc(30,-20){3}{0}
	\ArrowLine(30,-20)(60,-10)
	\ArrowLine(60,-30)(30,-20)
	\Photon(30,20)(30,7){2}{3}
	\Photon(30,-20)(30,-7){2}{3}
	\ArrowArc(30,0)(7,90,-90)
	\ArrowArc(30,0)(7,270,90)
\end{picture} 
\rule[-30pt]{0mm}{0mm}
	\ \sim \ \ 
	i\M_{0}
	\int_{l_1, l_2}
	\ 
	\bigl(\J_{1} \cdot \J_{2}\bigr)
	\ \ 
	e^2
	\frac{l_1^2}{l_2^2} 
	\mbox{Tr} \biggl(
	\frac{1}{\not l_1}\frac{1}{\not l_2-\not l_1}\biggr).
\ee
The loop integration over the fermion loop and over the
non-factorizable currents does not factorize.

As we have seen, two-loop non-factorizable corrections receive
contributions from two different sources: from additional interactions
with the decay products and from the corrections to the propagation
of the soft exchange particles.  The two contributions given by
\Eqns{feyn/2loop/nf}{feyn/2loop/nf_x} respectively are separately
gauge invariant and are distinguished in particular by a different
effective coupling constant. \Eqn{feyn/2loop/nf} originates solely
from additional interactions of photons with the decay products. This
term contains two non-factorizable currents for each decay and is
proportional to $\alpha^2$.  \Eqn{feyn/2loop/nf_x} on the other hand
receives contributions from propagator corrections as well as
interactions with the decay products. This term contains only one
non-factorizable current and is proportional to $\alpha^2 N_f$, where
$N_f$ is the number of fermions in the loop.

In summary, we separated the factorizable, $\sim \alpha_{\sss{fact}}$,
and non-factorizable corrections, $\sim \alpha_{\sss{nf}}$.  The
non-factorizable corrections are further separated into corrections
due to the interaction of the photons with the (dipole of the) decay
products, $\sim\alpha_{\sss{dipole}}$, and corrections due to the
propagation of the soft photons, $\sim \alpha_{\sss{prop}}$. Thus,
the complete two-loop corrections can be written schematically as
\be
 \alpha^2 = \alpha^2_{\sss{fact}}+\alpha_{\sss{fact}}\alpha_{\sss{nf}}
	+\alpha^2_{\sss{nf}};
	\qquad \qquad
 \alpha^2_{\sss{nf}}=\alpha_{\sss{dipole}}^2+
     (\alpha_{\sss{dipole}}\alpha_{\sss{prop}}).
\ee
The parenthesis in the last term indicates that the two effects do not
factorize.  The split up is gauge invariant and each contribution has
different physical properties.

\subsubsection{N-loop order}

The procedure described above generalizes to the separation of
$N$-loop correction.  The complete  $\alpha^N$ correction
consists of interferences between factorizable and non-factorizable
corrections, factorized with respect to each other.  Schematically
\be
\label{feyn/Nloop/full_struct}
	\alpha^N = \sum\limits_{i=0}^{N}
	\alpha^{N-i}_{\sss{fact}}\alpha_{\sss{nf}}^{i}.
\ee
Non-factorizable corrections can be due to the interaction of soft
photons with the dipoles of the decay products or due to the
correction to the propagation of the soft photons
\be 
\label{feyn/Nloop/nf_struct}
	\alpha^i_{\sss{nf}}
	=
	\sum\limits_{j=1}^{i}
	(\alpha_{\sss{dipole}}^{j}\alpha_{\sss{prop}}^{i-j}).
\ee
A useful illustration is given by the four-loop non-factorizable
diagrams shown in Fig.~\ref{fig:nonfactQED}.  The first diagram,
Fig.~\ref{fig:nonfactQED}(a), contains four dipole interactions and no
propagation corrections. The second diagram,
Fig.~\ref{fig:nonfactQED}(b), has three dipole interactions and a
first-order propagation correction. Finally, the last diagram,
Fig.~\ref{fig:nonfactQED}(e), has only one dipole interaction and a
three loop propagator correction.

\begin{figure}[ht]
\bigskip
\bigskip
\bigskip
\bigskip
\begin{center}
\begin{picture}(60,30)(0,-30) 
	\GCirc(0,0){3}{0}
	\DashLine(0,0)(30,20){3}
	\DashLine(0,0)(30,-20){3}
	\GCirc(30,20){3}{0}
	\ArrowLine(30,20)(60,30)
	\ArrowLine(60,10)(30,20)
	\GCirc(30,-20){3}{0}
	\ArrowLine(30,-20)(60,-10)
	\ArrowLine(60,-30)(30,-20)
	\PhotonArc(60,0)(36.05,150,210){1.5}{6}	
	\PhotonArc(0,0)(36.05,-30,30){1.5}{6}	
	\PhotonArc(40,0)(22.36,120,240){1.5}{7}	
	\PhotonArc(20,0)(22.36,-60,60){1.5}{7}
        \put(0,-30){(a)}
\end{picture}
\phantom{XX}
\begin{picture}(60,30)(0,-30) 
	\GCirc(0,0){3}{0}
	\DashLine(0,0)(30,20){3}
	\DashLine(0,0)(30,-20){3}
	\GCirc(30,20){3}{0}
	\ArrowLine(30,20)(60,30)
	\ArrowLine(60,10)(30,20)
	\GCirc(30,-20){3}{0}
	\ArrowLine(30,-20)(60,-10)
	\ArrowLine(60,-30)(30,-20)
	\PhotonArc(40,0)(22.36,120,240){1.5}{7}	
	\Photon(30,20)(30,7){1.5}{2}	
	\Photon(30,-7)(30,-20){1.5}{2}	
	\PhotonArc(20,0)(22.36,-60,60){1.5}{7}
	\ArrowArc(30,0)(7,90,-90)
	\ArrowArc(30,0)(7,270,90)
        \put(0,-30){(b)}
\end{picture}
\phantom{XX}
\begin{picture}(60,30)(0,-30) 
	\GCirc(0,0){3}{0}
	\DashLine(0,0)(30,20){3}
	\DashLine(0,0)(30,-20){3}
	\GCirc(30,20){3}{0}
	\ArrowLine(30,20)(60,30)
	\ArrowLine(60,10)(30,20)
	\GCirc(30,-20){3}{0}
	\ArrowLine(30,-20)(60,-10)
	\ArrowLine(60,-30)(30,-20)
	\Photon(30,20)(20,7){1.5}{3}	
	\Photon(30,20)(40,7){1.5}{3}	
	\Photon(20,-7)(30,-20){1.5}{3}
	\Photon(40,-7)(30,-20){1.5}{3}
	\ArrowLine(20,7)(40,7)
	\ArrowLine(40,7)(40,-7)
	\ArrowLine(40,-7)(20,-7)
	\ArrowLine(20,-7)(20,7)	
        \put(0,-30){(c)}
\end{picture}
\phantom{XX}
\begin{picture}(60,30)(0,-30) 
	\GCirc(0,0){3}{0}
	\DashLine(0,0)(30,20){3}
	\DashLine(0,0)(30,-20){3}
	\GCirc(30,20){3}{0}
	\ArrowLine(30,20)(60,30)
	\ArrowLine(60,10)(30,20)
	\GCirc(30,-20){3}{0}
	\ArrowLine(30,-20)(60,-10)
	\ArrowLine(60,-30)(30,-20)
	\Photon(20,7)(30,20){1.5}{3}	
	\Photon(40,7)(30,20){1.5}{3}
	\Photon(20,-7)(30,-20){1.5}{3}	
	\Photon(40,-7)(30,-20){1.5}{3}	
	\ArrowArc(20,0)(7,90,-90)
	\ArrowArc(20,0)(7,270,90)
	\ArrowArc(40,0)(7,90,-90)
	\ArrowArc(40,0)(7,270,90)
        \put(0,-30){(d)}
\end{picture}
\phantom{XX}
\begin{picture}(60,30)(0,-30) 
	\GCirc(0,0){3}{0}
	\DashLine(0,0)(30,20){3}
	\DashLine(0,0)(30,-20){3}
	\GCirc(30,20){3}{0}
	\ArrowLine(30,20)(60,30)
	\ArrowLine(60,10)(30,20)
	\GCirc(30,-20){3}{0}
	\ArrowLine(30,-20)(60,-10)
	\ArrowLine(60,-30)(30,-20)
	\Photon(30,10)(30,20){1.5}{2}	
	\Photon(30,-10)(30,-20){1.5}{2}	
	\ArrowArc(30,0)(10,90,-90)
	\ArrowArc(30,0)(10,270,90)
	\Photon(37.07,7.07)(22.929,-7.07){1.5}{5}	
	\Photon(37.07,-7.07)(22.929,7.07){1.5}{5}	
        \put(0,-30){(e)}
\end{picture}
 \ccaption{}{Four-loop diagrams contributing to non-factorizable
  corrections. The diagrams are proportional to (a)~$\sim\alpha_{\rm
  dipole}^4$, (b)~$\sim\alpha_{\rm dipole}^3\alpha_{\rm prop}$,
  (c)~$\sim\alpha_{\rm dipole}^2\alpha_{\rm prop}^2$,
  (d)~$\sim\alpha_{\rm dipole}^2 \alpha_{\rm prop}^2$ and
  (e)~$\sim\alpha_{\rm dipole} \alpha_{\rm
  prop}^3$. \label{fig:nonfactQED}} \end{center}
\end{figure}

At $N$-loop the non-factorizable correction arising solely from
dipole interactions, $\sim\alpha_{\sss{dipole}}^{N}$, can be written
in terms of the non-factorizable currents analogously to
\Eqns{feyn/1loop/nf}{feyn/2loop/nf}. They are given by
\be 
\label{feyn/Nloop/nf}
	i\M_{0} \ 
         \frac{1}{N!} \biggl[ \ * \int\frac{d^{4}k}{(2\pi)^4 k^{2}} \ 
                \bigl[\J_{1}\J_{2}\bigr](k) \ \biggr]^{N}.
\ee 
The products of non-factorizable currents are defined according to 
\Eqns{feyn/prod_of_currents}{feyn/prod_of_currents_x}.
Examination of the integral, \Eqn{feyn/Nloop/nf}, confirms that also
at higher-loop level the leading contribution to the non-factorizable
correction is due to soft photon exchange, $\Omega\sim \Gamma M/E$.

\subsubsection{Charged unstable particle production}

In this subsection we repeat the procedure of separating different
effects for the case of charged unstable particle production. The
difference to the neutral unstable particle case is that now in
addition to the decay-decay interferences there will be
production-decay interferences.

Let us consider a toy model with a massive, charged, scalar, unstable
particle field, $\phi$. The scalar field decays into a pair of massless
fermions, $\psi$ and $\psi'$. One of the fermions, $\psi$, is charged
and thus couples to a $U(1)$ gauge field, $A_{\mu}$.  The other
fermion, $\psi'$, is neutral.  The Lagrangian of the model is given by
\begin{eqnarray}
	\LL 
&=&
	\phi^\dagger \bigl(\cov{p}{}^2-M^2\bigr) \phi 
	+ \bar{\psi} \not\! \cov{p}{} \psi
	+ \bar{\psi'} \not\! p \psi'
	- \frac{1}{4} F_{\mu\nu}F^{\mu\nu}
\nonumber \\
&+&
	\biggl(
	  \PP X (\phi^\dagger \phi)
	+  \phi^\dagger \DD (\bar{\psi'} \psi)
	+\mbox{h.c.}
	\biggr)
\label{feyn/model_charged}
\end{eqnarray}

The difference to \Eqn{feyn/model} is that now the field $\phi$ is
charged, thus $\cov{p}{}$ in the kinetic term of the $\phi$ field is
the covariant derivative. Again, the $\PP$ and $\DD$ terms are
responsible for the production from a neutral source $X$ and for the
decay of the unstable particles respectively. As in the neutral case,
we assume that these terms can be treated perturbatively.

In order to see how the factorization picture can be generalized to
the case of charged unstable particles we consider the process
\be
  X \to \phi^+(p_1) \phi^-(p_2) 
  \to \psi^+(k_1)\psi'(k_1') \psi^-(k_2) \psi'(k_2').
\ee
where $p_{1,2}=k_{1,2}+k_{1,2}'$ and $p_{1,2}^2-M^2\sim \Gamma M$.
For simplicity we will assume that there are no charged initial state
particles. 

At one-loop the complete non-factorizable corrections can be written
as~\cite{dpa-calc}
\be
	 \M_{\sss{nf}}^{\sss{virt}}
	 =
	 i\M_{0}
	 \int\frac{d^{4}l}{(2\pi)^{4}l^{2}}
	 \biggl[
	 \J_{0}\cdot\J_{1}+\J_{0}\cdot\J_{2}+\J_{1}\cdot\J_{2}
	 \biggr].
\ee
The currents are given by
\be
	 \J_{0}^{\mu}
	 =
	 e\Biggl[
	  \frac{p_{1}^{\mu}}{lp_{1}}
	 +\frac{p_{2}^{\mu}}{-lp_{2}}
	 \Biggr],
\ee
for photon emission from the production stage of the process and 
\be
	 \J_{1}^{\mu}
	  =  -\,e\Biggl[ \frac{p_{1}^{\mu}}{lp_{1}}
        	        - \frac{k_{1}^{\mu}}{lk_{1}}
        	\Biggr]\frac{D_{1}}{D_{1}+2lp_{1}},
	 \ \ \ \ \ \ 
	 \J_{2}^{\mu}
	 = -\,e\Biggl[ \frac{p_{2}^{\mu}}{-lp_{2}}
        	        -\frac{k_{2}^{\mu}}{-lk_{2}}
	        \Biggr]\frac{D_{2}}{D_{2}-2lp_{2}}
\ee
for photon emission from the decay stages of the process. The main
difference is that there are three interference terms now. Two terms
account for photons connecting the production and the two decay
sub-stages (production-decay interferences). The third term
corresponds to photons connecting the two decay sub-stages
(decay-decay interferences). The decay-decay interferences are similar
to those in the neutral unstable particle case.

At $N$-loop order the full correction will be a sum of products
of factorizable and non-factorizable corrections similar to
\Eqn{feyn/Nloop/full_struct}.  Non-factorizable corrections come
from two sources similarly to \Eqn{feyn/Nloop/nf_struct}.  One is the
interaction of the soft photons with the decay or production dipoles.
This contribution can be expressed in terms of the non-factorizable
currents and is given by a generalization of \Eqn{feyn/Nloop/nf}. 
\be 
\label{feyn/Nloop/nf_x}
  i\M_{0} \ 
  \frac{1}{N!} \biggl[ \ * \int\frac{d^{4}l}{(2\pi)^4 l^{2}} \ 
   \bigl[
  \J_{0}\cdot\J_{1}+\J_{0}\cdot\J_{2}+\J_{1}\cdot\J_{2}\bigr](l) 
\ \biggr]^{N}.
\ee 
The second source for $N$-th order non-factorizable corrections is
photon propagation corrections to lower-order non-factorizable
corrections. Finally, factorizable corrections will be corrections to
the two decay sub-stages and to the production sub-stage.

In summary, in the case of production of charged unstable particles
the general picture still holds.  Indeed, at any order the full
correction comes from several physically different effects.  First of
all, factorizable and non-factorizable corrections can be separated.
Second, non-factorizable correction can be production-decay and
decay-decay interferences.  Each type of interference receives
corrections from the coupling of soft photons to production and decay
dipoles and from corrections to the soft photon propagation.

At this point let us comment on the possibility of a generalization of
this separation procedure to QCD.  We believe that in QCD the physical
picture should remain the same. At any order the corrections should
come from the physically different sources identified above and, thus,
one should be able to separate them.  It seems, however, that it will
be complicated to generalize the procedure we used so far.  This is
because the separation given above uses manipulations of the
non-factorizable currents.  In QCD the gauge structure is more
complicated due to the gluon self coupling. At higher order it will be
hard to maintain gauge invariance of separate contributions.  It would
be advantageous, therefore, to find a framework, which makes use of
the physical distinctions between various effects.  Such a framework
should be manifestly gauge invariant.  In Sect.~\ref{sec:eff} we come
back to this question.


\subsection{Integrating over decay products}

In this subsection we study the mechanism that leads to the
suppression of non-factorizable corrections at high energies, $E\gg
M$.  The mechanism is at work for distributions integrated over the
momenta of the decay products keeping their invariant masses fixed.
High energy suppression arises because of the coupling of the soft
photons to the decay products.  Therefore we will concentrate on the
non-factorizable corrections coming from the interaction with the
decay dipoles, as defined in Sect.~\ref{sec:factorization}.  We will
not consider any further factorizable corrections nor non-factorizable
propagation corrections.  They are proportional to the coupling
constant without any extra high energy suppression.  This fact
illustrates that the effects we identified in
Sect.~\ref{sec:factorization} are indeed physically different. Also,
we do not discuss high energy behavior of production-decay
interferences that are present in the charged case. The suppression
mechanism works differently in this case and we leave this issue for
the future.  Thus, the estimates that will be presented in this
subsection are valid for decay-decay non-factorizable interferences
coming from the interaction with decay dipoles. In
Sect.~\ref{sec:factorization} this was shown to be a well defined
gauge invariant subset of all corrections.

\subsubsection{Neutral Unstable Particles}

The first process we will consider is a neutral, unstable, scalar
particle with momentum $p$ decaying into a pair of massless, charged
fermions with momenta $k$ and $k'$ respectively.  The Lagrangian of
the model is given in \Eqn{feyn/model}.  We will estimate the $N$-th
order virtual non-factorizable correction due to exchange of
$N$-photons with momenta $l_1,\ldots,l_N$. As a first step we will
integrate the decay part of the Born matrix element squared multiplied
by $N$ non-factorizable currents over $k$ and $k'$, keeping $p$
fixed. According to  \Eqn{feyn/Nloop/nf} the relevant integral is
\be
\label{dec/int}
	I^{\alpha_1\ldots\alpha_N}
	=
	\int d^4k \ d^4k'  \ 
	\delta(k^2) \ \delta(k^{\prime 2}) \ 
	\delta^4(p-k-k') 
	\times |\DD|^2 \times
	\biggl(\frac{k}{kl_1}-\frac{k'}{k'l_1}\biggr)^{\alpha_1}
	\ldots
	\biggl(\frac{k}{kl_N}-\frac{k'}{k'l_N}\biggr)^{\alpha_N}.
\ee
This tensor corresponds to the decay part of one of the particles.
$|\DD|^2$ is the Born matrix element squared. The indices
$\alpha_1\ldots\alpha_N$ will be contracted with the indices of the
non-factorizable current from the other unstable particle.

In the center-of-mass frame at high energy the components of momenta in
the formula above can be estimated as follows:
\be
\label{dec/momenta}
	p\sim k \sim k'\sim E,
	\ \ \ 
	l_i\sim \Gamma M/E.
\ee 
The latter estimate follows from the fact that only soft photons are
relevant for non-factorizable corrections.  Also
$(kl_i)\sim(k'l_i)\sim(pl_i)\sim \Gamma M$.  Since ultra-relativistic
particle decay forward $k_\mu$ and $k'_\mu$ are almost collinear to
$p_\mu$ and to each other. This will lead to a suppression of the
non-factorizable currents in \Eqn{dec/int}. In order to make this
suppression explicit, it is convenient to change variables
\be
	k^\mu = \frac{1}{2}\bigl(p^\mu+\Delta^\mu\bigr),
	\ \ \ 
	k^{\prime \mu} = \frac{1}{2}\bigl(p^\mu-\Delta^\mu\bigr),
\ee
where $\Delta\sim M$. The phase-space integration over the final state
particle momenta is given by
\be
\label{dec/delta_int}
	\int d^4k \ d^4k'  \ 
	\delta(k^2) \ \delta(k^{\prime 2}) \ 
	\delta^4(p-k-k') 
	=
	\int d^4\!\Delta \ \delta(M^2+\Delta^2) \ \delta(2p\Delta)
	\equiv
	\int_{\Delta}
\ee
which allows to rewrite \Eqn{dec/int} as
\be
\label{dec/int_1}
	I^{\alpha_1\ldots\alpha_N}
	=
	\int_\Delta |\DD|^2
	\ 
	\biggl(\frac{k}{kl_1}-\frac{k'}{k'l_1}\biggr)^{\alpha_1}
	\ldots
	\biggl(\frac{k}{kl_N}-\frac{k'}{k'l_N}\biggr)^{\alpha_N}.
\ee

In terms of the new variables, the non-factorizable currents depend on
$p$ and $\Delta$.  They can be expanded in $M/E$
\be
	\biggl(\frac{k}{kl}-\frac{k'}{k'l}\biggr)^{\alpha}
	\ = \ 
	\Delta_{\alpha'} \ \cdot \ \frac{2}{lp} \ 
	\biggl(g^{\alpha'\alpha}-\frac{l^{\alpha'}p^\alpha}{lp}\biggr)
	\ + \ \OO(\Delta^3) 
	\ \sim \  
	\frac{1}{\Gamma}.
\label{current_exp}
\ee
There will be only odd powers of $\Delta$ in the expansion because the
exact non-factorizable current is anti-symmetric under
$k\leftrightarrow k'$.

We are now ready to estimate the size of $N$-non-factorizable currents
averaged over the momenta of decay products,
$I^{\alpha_1\ldots\alpha_N}$, in terms of $E$, $M$, and $\Gamma$. We
want to compare this to the leading contribution of the Born matrix
element squared, corresponding to $N=0$. This is because our goal
at the end is to estimate non-factorizable corrections with respect to
the Born cross section.

The estimate depends on the parity of $N$. If $N$ is odd the product
of the currents contains an odd number of $\Delta_{\alpha_i}$. Upon
integration these terms will vanish. For even $N$
\Eqns{dec/int_1}{current_exp} lead to the following estimate:
\be
\label{dec/est}
	I^{\alpha_1\ldots\alpha_N}
	\sim
	\left\{
	\begin{array}{l}
	0,  \ \ \ \mbox{for $N$ \ odd},\\
	\mbox{Born} \cdot \Gamma^{-N}, \ \ \  \mbox{for $N$ \ even}.
	\end{array}
	\right. 
\ee
In \Eqn{dec/est} we denote the leading contribution to the Born matrix
element squared by ``Born''.  Of course, the equation above does
not account for the Lorentz structure. However, the information it
provides is sufficient to estimate the non-factorizable correction.

For the purpose of illustration let us make this estimates more
precise by taking into account the Lorentz structure. As an example,
we consider the simplest case $N=2$:
\be
	I^{\alpha_1\alpha_2}
	\sim
	|\DD|^2
	\frac{p^2}{(pl_1)(pl_2)}
	\biggl[g^{\alpha_1\alpha_2}
	+\frac{(l_1l_2)}{(pl_1)(pl_2)}p^{\alpha_1} p^{\alpha_2}
	-\frac{1}{(pl_1)}l_1^{\alpha_2}p^{\alpha_1}
	-\frac{1}{(pl_2)}l_2^{\alpha_1}p^{\alpha_2}\biggr].
\ee
The first factor is the leading contribution to the Born term while the
second factor is the leading contribution from the non-factorizable
currents.  Power counting in the second factor gives $\sim
1/\Gamma^2$, in agreement with \Eqn{dec/est}.

The estimate given in \Eqn{dec/est} can be summarized in the following
way: using the scaling rules given in \Eqn{dec/momenta} the integrals
over the decay product momenta, \Eqns{dec/int}{dec/int_1}, can naively
be estimated to be of the order $\OO(\bigl(E/M\cdot 1/\Gamma\bigr)^N)$
relative to the Born term.  At high energy, however, $k$ is collinear to
$k'$, which introduces an extra suppression of $\OO(M^N/E^N)$ for
even $N$.  This results in the final estimate quoted above:
$\OO(\Gamma^{-N})$ relative to the Born term.

Finally we are in a position to estimate the non-factorizable
correction itself.  At $N$-th order it is given by
\Eqn{feyn/Nloop/nf} and involves $N$-integrations over the momenta
of virtual particles, $N$ propagators of soft exchange particles and
interference of $2N$ non-factorizable currents, $N$ coming from each
unstable particle.  We assume that one integrates over the momenta of
decay products. This allows us to use the estimates made in the
previous subsection. Denoting the non-factorizable correction relative
to the Born term by $\delta_{\sss{nf}}$, we find for even $N$
\be
	\delta_{\sss{nf}}
	\sim
	\biggl(\frac{(\Gamma M/E)^4}{(\Gamma M/E)^2}\biggr)^N
	\ \cdot \ 
	\biggl(\frac{1}{\Gamma}\biggr)^N
	\ \cdot \ 
	\biggl(\frac{1}{\Gamma}\biggr)^N
	\sim
	\biggl(\frac{M}{E}\biggr)^{2N},
	\ \ \ \ \ \ 
	N\ \mbox{even}.
\label{feyn:nfneutraleven}
\ee
The first factor comes from the loop integration measure and $N$
virtual particle propagators whereas the second and the third factors
are due to the $2N$ non-factorizable currents. The non-factorizable
correction for odd $N$
\be
	\delta_{\sss{nf}}
	= 0,
	\ \ \ \ \ \ 
	N\ \mbox{odd},
\label{feyn:nfneutralodd}
\ee
because of the symmetry mentioned above.

\subsubsection{Charged Unstable Particles}

The estimates made in the previous subsection can be generalized to
the case where the unstable particle is charged. We will consider a
process where the unstable particle, with momentum $p^\mu$, decays
into a charged and neutral fermion with momentum $k^\mu$ and
$k^{\prime \mu}$ respectively. The Lagrangian of the model is given in
\Eqn{feyn/model_charged}. The Born matrix element squared, multiplied
by $N$ non-factorizable currents and integrated over the decay product
momenta now reads
\be
\label{dec/int_1_x}
	I^{\alpha_1\ldots\alpha_N}
	=
	\int_\Delta |\DD|^2
	\cdot
	\biggl(\frac{p}{pl_1}-\frac{k}{kl_1}\biggr)^{\alpha_1}
	\ldots
	\biggl(\frac{p}{pl_N}-\frac{k}{kl_N}\biggr)^{\alpha_N}.
\ee
As in the neutral case we change variables and define $\Delta^{\mu}$
through the relation $k=(p+\Delta)/2$ with $\Delta\sim M$. Expressing
the non-factorizable currents in the new variables and expanding in
$M/E$ results in
\bea
	\biggl(\frac{p}{pl}-\frac{k}{kl}\biggr)^{\alpha}
	&=& 
	- \Delta_{\alpha'} \ \cdot \ \frac{1}{lp} \ 
	\biggl(g^{\alpha'\alpha}-\frac{l^{\alpha'}p^\alpha}{lp}\biggr)
	\ + \  
	\Delta_{\alpha'} \Delta_{\beta'} \ \cdot \ \frac{l^{\beta'}}{(lp)^2} \ 
	\biggl(g^{\alpha'\alpha}-\frac{l^{\alpha'}p^\alpha}{lp}\biggr)
	\ + \ 
	\OO(\Delta^3) 
\nonumber \\
& \sim &
	\frac{1}{\Gamma} \biggl(1  \ + \ \frac{M}{E} \ + \ \ldots \biggr).
\eea
The main difference compared to the case of neutral, unstable
particles is that in the charged case there are odd powers of $\Delta$
in the expansion. This is because the exact non-factorizable current
is not anti-symmetric under $k\leftrightarrow k'$ anymore.  As a
consequence non-factorizable correction will not vanish any longer for
odd $N$.  Instead they will get an extra $M/E$ suppression. The reason
for this additional suppression is that one of the non-factorizable
currents will have to be expanded to second order in $\Delta$. For
even $N$ the estimates are the same as in the neutral case. Repeating
the estimates that lead to \Eqn{dec/est} one obtains
\be
\label{dec/est_x}
	I^{\alpha_1\ldots\alpha_N}
	\sim
	\left\{
	\begin{array}{l}
	\mbox{Born} \cdot \Gamma^{-N} \cdot M/E,  \ \ \ \mbox{$N$ \ odd},\\
	\mbox{Born} \cdot \Gamma^{-N}, \ \ \  \mbox{$N$ \ even}
	\end{array}
	\right. 
\ee
where ``Born'' again denotes the leading contribution to the Born
matrix element squared.

As in the neutral case it is possible to get the Lorentz structure.
Let us consider, for example, the case $N=1$:
\be
	I^{\alpha}
	\sim
	|\DD|^2
	\ \cdot \ 
	\frac{p^2}{(pl)^2}
	\biggl[l^\alpha-\frac{l^2}{(lp)} p^\alpha\biggr].
\ee
The first factor is the leading contribution to the Born term whereas the
second factor is the leading contribution from the non-factorizable
currents.  Power counting in the second factor gives $\sim 1/\Gamma
\cdot M/E$.

Using \Eqn{dec/est_x} we can now obtain the estimate of the
decay-decay non-factorizable correction in the charged case. We obtain
\bea
	\delta_{\sss{nf}}
	&\sim&
	\biggl(\frac{M}{E}\biggr)^{2N},
	\qquad
	N\ \mbox{even},
\label{feyn:est_even}
\\
	\delta_{\sss{nf}}
	&\sim&
	\biggl(\frac{M}{E}\biggr)^{2N+2},
	\qquad
	N\ \mbox{odd},
\label{feyn:est_odd}
\eea
These estimates can be summarized as follows: the non-factorizable
correction with $2N$ decay currents scales as $E^{-2N}$ if $N$ is
even.  Because of symmetry reasons there is an extra suppression of
$1/E$ for each of the two phase-space integrations over the angles if
$N$ is odd.  It should be stressed again that this estimate is
incomplete. The reason is that we did not consider production-decay
interferences even though they are present in the charged case.


\section{Effective Field Theory description}
\label{sec:eff}

The estimates of non-factorizable corrections presented in the
previous subsection seem to be quite technical. They show different
energy scaling behavior depending on the process, particles
participating in the process, type of interaction through which they
interact and many other factors.  Moreover, the very definition of
what we mean by non-factorizable corrections rests on the separation
of factorizable and non-factorizable corrections discussed in
Sect.~\ref{sec:factorization}. There we discussed just one specific
example (out of many) and showed how such a separation can be achieved
in QED.

Despite all the different ways the non-factorizable corrections
manifest themselves, all the estimates rely on one simple
observation. In all processes involving the production of unstable
particles close to resonance there are always two scales present: a
hard scale, $\Lambda_1\sim E$, and a soft scale, $\Lambda_2\sim\Gamma
\cdot M/E$. If $E^2\gg M\Gamma$ there is a hierarchy between these two
scales, $\Lambda_1\gg \Lambda_2$.  At the hard scale, $\Lambda_1$, the
process can be described in terms of interacting stable particles. The
fact that the particles are unstable is not very important and can be
neglected.  At the soft scale, $\Lambda_2$, the instability of the
particles becomes important and cannot be neglected any longer. If
one's objective is to study effects due to the instability then the
problem can be simplified by separating the two contributions and
studying them separately.

This separation of scales is most naturally reflected in an effective
field theory framework. In this framework one starts with identifying
all the relevant degrees of freedom in the underlying theory. They are
determined by the scales present in the process. These scales can come
from two sources. Either they come from the underlying theory itself
(masses of heavy particles, etc.) or they are introduced through
kinematical constraints (CMS energy, etc.).  Without kinematical
scales the mode contribution is determined by the typical excitation
energy.  However, in the presence of kinematically induced scales the
excitation energy alone is not sufficient to determine the mode
contribution.  Additional quantum numbers are needed to describe the
mode contributions uniquely.%
\footnote{In the case of heavy quark pair production close to
 threshold, for example, the kinematical constraints are such that
 particles are produced with a small velocity, $v$ (see \cite{nrqcd,
 beneke-smirnov}). There are three modes distinguished by energy with
 four-momenta: $(\sim m,\sim m)$, $(\sim mv,\sim mv)$, $(\sim
 mv^2,\sim mv^2)$.  This is not the complete list, however. One needs
 an additional quantum number --- three-momentum --- to identify an
 additional mode with four-momentum $(\sim mv,\sim mv^2)$. }
Once the relevant modes have been identified we can integrate out the
modes that do not show up as external particles.

Let us recall how the procedure of constructing an effective theory works
in the simpler case without kinematical constraints. The first step is to
integrate out hard momenta. This results in a theory where only soft
degrees of freedom are dynamical. The interactions of the soft degrees of
freedom are described by an effective Lagrangian
\be 
\LL_{\sss{eff}} =
\sum\limits_{n} \frac{c_n}{\Lambda^{n}} \ \OO_{n}(\mbox{soft fields}).
\ee 
Each operator, $\OO_n$, is multiplied by some power of a scale,
$\Lambda$, and a dimensionless Wilson coefficient, $c_n$, which is a
series in the coupling constant of the original theory $c_{n} = \sum
c_{n,i}\alpha^{i}$.  The contributions of the hard modes are included
in the Wilson coefficients. These coefficients are obtained by
matching the effective theory to the original theory. Since there is a
hierarchy of scales the operators $\OO_n$ are local and $\Lambda$ is a
typical excitation energy of the hard modes.

Let us now investigate how this picture applies to processes with
unstable particles produced close to resonance.  This induces a
kinematical constraint and consequently the distinction between the
various modes is more subtle.  First of all, modes are distinguished
by their momentum, which can either be hard, $P\sim\Lambda_1$, or
soft, $p\sim\Lambda_2$. By the statement $P\sim\Lambda_1$ we mean that
at least one component of the momentum is of the order $\Lambda_1$ and
none is much larger. However, the fact that there is a kinematical
constraint means that an additional quantum number is necessary to
distinguish different modes.  This quantum number is off-shellness or
virtuality of the particles, $D$, defined as
\be
D\equiv P^2-m^2, 
\ee 
where $P$ is the total momentum and $m$ the mass of the particle.  Indeed,
if the particle is close to resonance, $P^2\sim m^2$, it can happen that a
particle has large momentum, $P\sim\Lambda_1$, but its virtuality is still
small, or even zero.  In this configuration observables are sensitive to
small (soft) changes of momentum, $P\to P\pm p$, $p\sim\Lambda_2$.  The
contribution of such modes depends on two momenta: the hard on-shell
momentum, $P$, and soft momentum, $p$.  The total momentum of such a mode
is $P+p\sim\Lambda_1$, and the virtuality is $D\sim \Lambda_1\Lambda_2$.

This makes it clear that in order to account correctly for all the
relevant modes we will have to distinguish between three types of
momenta:
\begin{itemize}
\item   off-shell hard momenta, $P\sim\Lambda_1$, $P^2-m^2\sim\Lambda_1^2$.
\item   on-shell hard momenta, $P_m\sim\Lambda_1$, $P_m^2=m^2$ 
\item   soft momenta, $p\sim\Lambda_2$.
\end{itemize}
In general, the virtuality can take four values: 
0, $\Lambda_2^2$, $\Lambda_1\Lambda_2$, and $\Lambda_1^2$.
Consequently, there are the following types of modes in the problem:
\begin{itemize}
\item   Hard modes: $\Phi(P)$, where $P\sim \Lambda_1$, $D\sim
  \Lambda_1^2$.
\item   Resonant modes $\Phi(P_m,p)$, where $P_m\sim\Lambda_1$, $p\sim
  \Lambda_2$, $D\sim\Lambda_1\Lambda_2$.
\item   Soft modes: $\Phi(p)$, where $p\sim\Lambda_2$, $D\sim\Lambda_2^2$.
\item   External modes: $\Phi(P_m)$, where $P_m\sim\Lambda_1$, $D=0$.
\end{itemize}
All fields together with their respective energy and virtuality scales
are listed in Fig.~\ref{fig:scales}. 

\begin{figure}[ht]
\bigskip
\bigskip
\begin{center}
 \begin{picture}(40,80)(0,0) 
   \Line(20,-5)(20,80)\Line(20,80)(15,75)\Line(20,80)(25,75)
   \put(0,77){\small $P$}
   \Line(15,0)(25,0)       \put(2,-3){$0$}
   \Line(15,30)(25,30)     \put(0,27){$\Lambda_2$}         
   \put(30,27){$\Phi(P_m,p), \Phi(p)$}
   \Line(15,60)(25,60)     \put(0,57){$\Lambda_1$}         \
   \put(30,57){$\Phi(P_m,p), \Phi(P), \Phi(P_m)$}
\end{picture}
\phantom{XXXXXXXXXXXXXXX}
 \begin{picture}(40,80)(0,0) 
   \Line(20,-5)(20,80)\Line(20,80)(15,75)\Line(20,80)(25,75)
   \put(0,77){\small$D$}
   \Line(15,0)(25,0)\put(0,-3){$0$}  \put(30,-3){$\Phi(P_m)$}
   \Line(15,20)(25,20)  
   \put(-5,17){$\Lambda_2^2$} \put(30,17){$\Phi(p)$}
   \Line(15,40)(25,40)  
   \put(-15,37){$\Lambda_1\Lambda_2$}  \put(30,37){$\Phi(P_m,p)$}
   \Line(15,60)(25,60)     
   \put(-5,57){$\Lambda_1^2$}      \put(30,57){$\Phi(P)$}
\end{picture}
 \ccaption{}{Relevant energy-momentum and virtuality
                scales with a list of corresponding fields
                contributing at each scale. \label{fig:scales}}
  \end{center}
\end{figure}

Let us now consider what happens if we start to integrate out fields
in the case where kinematical scales are present. It is possible that
fields that are integrated out and fields that are left in the
Lagrangian -- being distinguished by the additional quantum number --
have the same typical energy.  In this case the operators $\OO_n$ can
be nonlocal. The scale $\Lambda$ is determined by the quantum number,
which distinguishes the modes.

Applying this strategy to processes with unstable particles one first
integrates out the hard modes, $\Phi(P)$.  The resulting effective theory
describes instability effects, such as propagation of unstable particles
close to resonance, $\Phi(P_m,p)$, their decay into on-shell particles,
$\Phi(P_m)$, and interactions between them due to soft exchanges,
$\Phi(p)$.  The effective Lagrangian contains nonlocal operators.  The
main point is that in this way we achieve a separation of factorizable and
non-factorizable effects. As we will see, the factorizable corrections
correspond to the hard effects encoded in the Wilson coefficients whereas
all non-factorizable corrections are described by interactions of the
still dynamical soft modes. Since this separation is based on a physically
relevant hierarchy of scales it is possible to generalize it to more
complicated cases. The usual way of separating factorizable and
non-factorizable corrections is based on {\it ad hoc} manipulations of
various contributions. These manipulations become increasingly tedious for
more complicated cases.

As a next step one can integrate out the resonant modes, $\Phi(P_m,p)$.
This results in a new effective theory that contains nonlocal interactions
between external particles, $\Phi(P_m)$, coupled in the production and decay
points. It also contains interactions between production and decay points
due to soft exchanges, $\Phi(p)$ (non-factorizable corrections).  Whereas
integrating out hard modes provides a separation between factorizable and
non-factorizable corrections, integrating out resonant modes provides a
separation of different types of non-factorizable corrections
(production-decay, decay-decay). 

The last step is to integrate out the soft modes, $\Phi(p)$. This will
lead to the $S$-matrix defined in terms of external particles,
$\Phi(P_m)$.

\subsection{Toy model}
\label{sec:eff/model}

Let us now carry out this programme in some more detail for the toy
model specified in \Eqn{feyn/model}.  The list of relevant modes is
the following:
\begin{itemize}
\item   Hard modes: $\phi(P)$, $\psi(P)$, $A_{\mu}(P)$.
\item   Resonant modes: $\phi(P_M,p)$, $\psi(P_0,p)$.
\item   Soft modes: $\psi(p)$, $A_{\mu}(p)$.
\item   External modes: $\psi(P_0)$.
\end{itemize}
It should be understood that the $A_{\mu}(P)$ field is subject to gauge
transformations at the hard scale whereas $A_{\mu}(p)$ is subject to gauge
transformations at the soft scale.  Thus, for example, the covariant
derivative acting on $\psi(P)$ contains $A_{\mu}(P)$ and the covariant
derivative acting on $\psi(P_0,p)$ contains $A_{\mu}(p)$.

The leading diagram for the process \Eqn{toy-process} is shown in
Fig.~\ref{fig:eff/born}(a) where the dashed line denotes a resonant
field $\phi(P_M,p)$. Neglecting an overall coupling constant, this
diagram is of the order $(\Lo \Lt)^{-2}$.

\subsection{Integrating out hard modes}
\label{sec:eff/hard}

The Lagrangian given in \Eqn{feyn/model} describes the full theory
where all modes are dynamical. The first step towards the effective
theory we are aiming at is to integrate out the hard modes $\phi(P)$,
$\psi(P)$, and $A_\mu(P)$. They all have momenta $P \sim \Lo$ and
virtuality $D\sim\Lo^2$. The resulting effective Lagrangian contains
fields $\phi(P_M,p)$ ($P_M\sim\Lo, P_M^2=M^2$), $\psi(P_0,p)$
($P_0\sim\Lo, P_0^2=0$), $\psi(p)$ and $A_\mu(p)$, with $p\sim\Lt$,
and the external field $\psi(P_0)$.  The Lagrangian of the effective
theory is determined through matching.  Thus, we require that up to a
certain order in the coupling constants and up to a certain order in
$\Lt/\Lo$ the effective theory coincides with the underlying
theory. At this point we restrict ourselves to contributions that are
not suppressed by powers of $\Lt/\Lo$.

\begin{figure}[ht]
\bigskip
\bigskip
\begin{center}
 \begin{picture}(80,60)(0,-30) 
        \GCirc(0,0){3}{0}
        \DashLine(0,0)(30,20){3}
        \DashLine(0,0)(30,-20){3}
        \GCirc(30,20){3}{0}
        \ArrowLine(30,20)(60,30)
        \ArrowLine(60,10)(30,20)
        \DashCArc(30,20)(35,-20,20){1}\put(70,20){\small $D_1$}
        \GCirc(30,-20){3}{0}
        \ArrowLine(30,-20)(60,-10)
        \ArrowLine(60,-30)(30,-20)      
        \DashCArc(30,-20)(35,-20,20){1}\put(70,-20){\small $D_2$}
        \put(0,-30){(a)}
\end{picture} 
\phantom{XXXXXXXXXX}
\begin{picture}(80,60)(0,-30) 
        \GCirc(0,0){3}{0}
        \DashLine(0,0)(30,20){3}
        \DashLine(0,0)(30,-20){3}
        \GCirc(30,20){3}{0}
        \ArrowLine(30,20)(60,30)
        \ArrowLine(60,-10)(30,20)
        \DashCArc(30,20)(35,-20,20){1}\put(70,20){\small $D_1$}
        \GCirc(30,-20){3}{0}
        \ArrowLine(60,10)(30,-20)
        \ArrowLine(30,-20)(60,-30) 
        \DashCArc(30,-20)(35,-20,20){1}\put(70,-20){\small $D_2$}
        \put(0,-30){(b)}
\end{picture}
\ccaption{}{Resonant (a) and background (b) Born contributions to pair
        production in the underlying theory. Resonant kinematics means
        that $D_1\sim D_2\sim \Lo\Lt$.  \label{fig:eff/born}} 
  \end{center}
\end{figure}

Let us start by integrating out $\phi(P)$. At leading order in $\alpha$,
we have to consider diagrams as shown in Fig.~\ref{fig:eff/born}.  The
diagram shown in Fig.~\ref{fig:eff/born}(a) cannot contribute to the
process \Eqn{toy-process} if the $\phi$-field is hard, since this would
violate the kinematic constraint \Eqn{kin-const}.  On the other hand, the
background process, Fig.~\ref{fig:eff/born}(b), does contribute to this
process, even if the $\phi$-fields are hard. Thus, integrating out
$\phi(P)$ results in an operator $X (\bar{\psi} \psi)(\bar{\psi} \psi)$ in
the effective Lagrangian. However, for each hard $\phi$-field we get a
suppression of $\Lt/\Lo$. Thus, in leading order in  $\Lt/\Lo$ there are
no new operators introduced by integrating out $\phi(P)$.

Consider now $\alpha$-corrections in the leading $\Lt/\Lo$
approximation.  As shown in Fig.~\ref{fig:eff/otherhard}(a), there are
diagrams involving hard $\phi$-fields that are not suppressed by
$\Lt/\Lo$. However, these diagrams do not result in new
operators in the effective Lagrangian.  Rather, their effect is
encoded in the (modified) coefficients of already existing
operators. In fact, Fig.~\ref{fig:eff/otherhard}(a) results in a
modification of $\PP X\phi\phi$. Thus, $\PP X\phi\phi \to c_\PP
\PP X\phi\phi$ where $ c_\PP = 1 + \OO(\alpha)$ is a Wilson coefficient.
Of course, there are also diagrams resulting in $\alpha$ corrections to
operators that are suppressed by $\Lt/\Lo$.  An example of a diagram that
gives rise to an $\alpha$ correction to the operator $X (\bar{\psi}
\psi)(\bar{\psi} \psi)$ is shown in Fig.~\ref{fig:eff/otherhard}(b).

\begin{figure}[ht]
\bigskip
\bigskip
\begin{center}
\begin{picture}(80,60)(0,-30) 
        \GCirc(0,0){3}{0}
        \DashLine(0,0)(20,20){3}
        \DashLine(0,0)(20,-20){3}
        \GCirc(20,20){1}{0}
        \GCirc(40,20){1}{0}
        \GCirc(20,-20){1}{0}
        \GCirc(40,-20){1}{0}
        \ArrowLine(20,20)(40,20)
        \ArrowLine(40,20)(40,-20)
        \ArrowLine(40,-20)(20,-20)
        \ArrowLine(20,-20)(20,20)
        \DashLine(40,20)(60,20){3}
        \DashLine(40,-20)(60,-20){3}
        \ArrowLine(60,20)(80,30)
        \ArrowLine(80,10)(60,20)
        \GCirc(60,20){3}{0}
        \ArrowLine(60,-20)(80,-10)
        \ArrowLine(80,-30)(60,-20)      
        \GCirc(60,-20){3}{0}
        \put(0,-30){(a)}
\end{picture} 
\phantom{XXXXXXXXXX}
\begin{picture}(80,60)(0,-30) 
        \GCirc(0,0){3}{0}
        \DashLine(0,0)(30,20){3}
        \DashLine(0,0)(30,-20){3}
        \GCirc(30,20){1}{0}
        \ArrowLine(30,20)(60,30)
        \GCirc(30,-20){1}{0}
        \ArrowLine(30,-20)(30,20)
        \ArrowLine(60,-30)(30,-20)  
        \DashLine(30,0)(50,0){3}
        \ArrowLine(70,-10)(50,0)
        \ArrowLine(50,0)(70,10)  
        \GCirc(50,0){3}{0}
        \put(0,-30){(b)}
\end{picture}
\ccaption{}{Diagrams involving $\phi(P)$ that result in (a) $\alpha$
        corrections to  $\PP X \phi\phi$ and (b) $\alpha$ corrections to
        the operator $X (\bar{\psi} \psi)(\bar{\psi} \psi)$. In
        diagram (a) all loops are hard but the decaying $\phi$ fields
        are resonant. In diagram (b) all internal fields are
        hard. \label{fig:eff/otherhard} }
  \end{center}
\end{figure}

As a next step we turn to the $U(1)$ field. In order to integrate out
$A(P)$ we have to consider diagrams as shown in
Fig.~\ref{fig:eff/hard/corr}. For a hard photon, the diagram shown in
Fig.~\ref{fig:eff/hard/corr}(a) is suppressed by powers of $\Lt/\Lo$. This
diagram only gives a leading in $\Lt/\Lo$ contribution for a soft photon.
The diagram shown in Fig.~\ref{fig:eff/hard/corr}(b), however, is only
suppressed by a coupling constant $\alpha$, but not by powers of
$\Lt/\Lo$. Thus, it leads to an $\alpha$ correction to the contribution of
the $\phi\bar{\psi}\DD\psi$ operator. We take this into account by
multiplying this operator by a Wilson coefficient $c_\DD = 1+\OO(\alpha)$.
Since we are dealing with a neutral field $\phi$, the diagram shown in
Fig.~\ref{fig:eff/hard/corr}(c) does not contribute.  

\begin{figure}[ht]
\bigskip
\bigskip
\begin{center}
 \begin{picture}(80,60)(0,-30) 
        \GCirc(0,0){3}{0}
        \DashLine(0,0)(30,20){3}
        \DashLine(0,0)(30,-20){3}
        \GCirc(30,20){3}{0}
        \ArrowLine(30,20)(60,30)
        \ArrowLine(60,10)(30,20)
        \GCirc(30,-20){3}{0}
        \ArrowLine(30,-20)(60,-10)
        \ArrowLine(60,-30)(30,-20)      
        \Photon(45,15)(45,-15){2}{6}
        \put(0,-30){(a)}
\end{picture} 
\phantom{XXXXX}
 \begin{picture}(80,60)(0,-30) 
        \GCirc(0,0){3}{0}
        \DashLine(0,0)(30,20){3}
        \DashLine(0,0)(30,-20){3}
        \GCirc(30,20){3}{0}
        \ArrowLine(30,20)(60,30)
        \ArrowLine(60,10)(30,20)
        \GCirc(30,-20){3}{0}
        \ArrowLine(30,-20)(60,-10)
        \ArrowLine(60,-30)(30,-20)      
        \Photon(50,26.666)(50,13.333){2}{3}
        \put(0,-30){(b)}
\end{picture}
\phantom{XXXXX}
 \begin{picture}(80,60)(0,-30) 
        \GCirc(0,0){3}{0}
        \DashLine(0,0)(30,20){3}
        \DashLine(0,0)(30,-20){3}
        \GCirc(30,20){3}{0}
        \ArrowLine(30,20)(60,30)
        \ArrowLine(60,10)(30,20)
        \GCirc(30,-20){3}{0}
        \ArrowLine(30,-20)(60,-10)
        \ArrowLine(60,-30)(30,-20)      
        \PhotonArc(30,20)(15,215,-20){2}{5}
        \put(0,-30){(c)}
\end{picture}
\ccaption{}{One-loop radiative corrections in effective theory. (a) would
        be counted as non-factorizable and (b) as factorizable
        correction in the underlying theory. If the unstable particles
        are charged there are diagrams in the underlying theory that
        contain both factorizable and non-factorizable corrections. An
        example is given in (c). \label{fig:eff/hard/corr}}
        \end{center}
\end{figure}

The fields $\psi(P)$ enter through self energy insertions of
$\phi(P_M,p)$, thus they will modify the operators bilinear in
$\phi$. The corresponding operator leading in $\Lt/\Lo$ is $\phi
P_M^2\phi=\phi M^2\phi$. However, at leading order in $\alpha$ this
operator does not contribute. The next-to-leading order corrections in
$\alpha$ result in an operator $\phi iM\Gamma\phi$. This can be seen
as follows: the relative order of magnitude of a hard self energy
insertion is given by $\alpha \Lo^4 \Lo^{-2} (\Lo \Lt)^{-1} = \alpha
\Lo/\Lt$. The factor $\alpha \sim g^2$ comes from the coupling
constant, $\Lo^4$ from the phase-space integration, $\Lo^{-2}$ from
the hard fermion propagators and $(\Lo \Lt)^{-1}$ from the additional
$\phi(P_M,p)$ propagator. From the estimate above it is evident that
the self energy insertion is not suppressed. Therefore, these
contributions have to be taken into account to all orders. This leads
to the usual self energy resummation.

In summary, at leading order in $\Lt/\Lo$ the effective Lagrangian takes
the form
\be
\label{eff/hard}
        \LL_{\sss{eff}} =
         \frac{1}{2} \phi \bigl(2 P_M p+iM\Gamma\bigr) \phi 
        + \bar{\psi}(\not\!P_0\,+\!\not\!\cov{p}\,) \psi
        - \frac{1}{4} F_{\mu\nu}F^{\mu\nu}
        + c_\PP \phi X\, \PP(P_M) \phi 
        + c_\DD \phi\bar{\psi} \DD(P_M) \psi.
\ee
We should stress that the terms $\bar{\psi}\!\not\!\cov{p}\, \psi$ and
$F_{\mu\nu}F^{\mu\nu}$ in \Eqn{eff/hard} are not equivalent to the
ones in \Eqn{feyn/model}. In the underlying theory the fields $A(P)$
and $\psi(P)$ are dynamical whereas in \Eqn{eff/hard} they have been
integrated out. 

The effective Lagrangian given in \Eqn{eff/hard} is in fact completely
equivalent to the double pole approximation (DPA) \cite{dpa}. At Born
level this can be seen as follows: The production and decay operators,
$\phi\PP(P_M)\phi$ and $\phi\bar{\psi}\DD(P_M)\psi$, depend on the
on-shell momentum $P_M$. The kinetic operator of the unstable particle
depends also on the soft momentum $p$.  Consequently, the Born process
shown in Fig.~\ref{fig:eff/born}(a) is described by the Lagrangian
\Eqn{eff/hard} as a superposition of an on-shell production subprocess,
unstable particle propagation subprocess, and on-shell decay subprocess.
This picture is the same as that in DPA.

The equivalence of the effective theory given by the Lagrangian
\Eqn{eff/hard} and DPA can be extended to $\alpha$-corrections in the
leading $\Lt/\Lo$ approximation.  In DPA the important fact is that
$\alpha$-corrections are separated into a sum of factorizable and
non-factorizable corrections. The diagram shown in
Fig.~\ref{fig:eff/hard/corr}(a) is a non-factorizable correction. Only
soft gauge bosons give contributions to non-factorizable corrections in
DPA. On the other hand the diagram Fig.~\ref{fig:eff/hard/corr}(b) is a
factorizable correction in DPA.

Let us see how this happens in our effective theory framework.  There
are two sources of $\alpha$-corrections.  First, the Wilson
coefficients contain $\alpha$-corrections.  As we have seen above,
these correspond to corrections obtained by integrating out hard
modes.  Second, there are loop corrections in the effective theory due
to the exchange of soft gauge bosons. In the effective theory at
leading $\Lt/\Lo$ order gauge bosons couple to resonant particles in
the soft photon approximation. Thus, the evaluation of a diagram of
the type Fig.~\ref{fig:eff/hard/corr}(a) with a soft photon is the
same in the effective theory and the DPA of the underlying
theory. Furthermore, loop integrals in the effective theory have to be
carried out in dimensional regularization. A loop integral
corresponding to the diagram of Fig.~\ref{fig:eff/hard/corr}(b) with a
soft photon does not have a scale and, therefore, is zero in
dimensional regularization.

All of the above means that the complete factorizable corrections
correspond to the contributions due to the hard modes, calculated in
dimensional regularization. These contributions are included in the Wilson
coefficients of the effective theory. The non-factorizable corrections are
described by the dynamical degrees of freedom of the effective theory. In
this way we achieve a separation of factorizable and non-factorizable
effects based on the hierarchy of scales in the problem.  It is clear that
the effective theory description provides a separation between
factorizable and non-factorizable corrections also at higher orders in
$\alpha$.  This is the main advantage of the effective theory approach
compared to DPA.

In more complicated cases, for example when the unstable particles are
charged, the separation between factorizable and non-factorizable
corrections cannot be done on the basis of diagrams.  There are diagrams,
like the one shown in Fig.~\ref{fig:eff/hard/corr}(c), which contain both
factorizable and non-factorizable parts.  The separation is still
correctly reproduced within the effective theory framework.  In
Appendix~\ref{app:a} we show how the separation works for this example to
all orders in $\Lt/\Lo$.  In order to do that we use an expansion of loop
integrals in dimensional regularization that is equivalent to effective
theory calculations. 

The effective Lagrangian~\Eqn{eff/hard} contains only local operators.  On
general grounds, however, one can expect nonlocal operators to be present.
This is because the momentum of the fields integrated out, $P\sim\Lo$, is
of the same order as the momentum of the fields still present,
$P_M\sim\Lo$.  In order to understand this mismatch let us recall that in
our example hard modes do not introduce new operators at leading order in
$\Lt/\Lo$.  Thus, the resonant diagrams are described in terms of the same
operators in the original and effective theory. These operators are local.
As mentioned before, hard modes will introduce new operators that are
suppressed by $\Lt/\Lo$. Consequently, nonlocal operators will appear in
the effective Lagrangian only at next to leading order in $\Lt/\Lo$.

This brings us to the question of $\Lt/\Lo$ suppressed operators. As we
have shown, the effective Lagrangian \Eqn{eff/hard} contains leading
operators in $\Lt/\Lo$ and is completely equivalent to DPA.  The effective
field theory approach should enable us -- at least in principle -- to
construct $\Lt/\Lo$ suppressed operators order by order in $\Lt/\Lo$.
This would open the possibility to go beyond DPA and calculate $\Lt/\Lo$
corrections in a regular (in particular, gauge invariant) way. It is not
the aim of this paper to go beyond the leading in $\Lt/\Lo$ order.
However, before proceeding, we would like to discuss this point briefly.
In the construction of the effective Lagrangian we assumed the presence of
two widely separated scales. One of the scales, $\Lt$, was induced by the
kinematical constraints that the unstable particles are close to
resonance. If the unstable particles are far from resonance then there is
no extra scale, and the whole effective theory approach is not applicable.
This means that for different parts of the phase space one needs to
perform different calculations and switch between the two as one goes from
one phase-space region to the other. There is a question of accuracy and
applicability of the calculation in the transition region.  This question
always arises when effective theories are used for exclusive
distributions.  If one is interested in inclusive distributions, where one
integrates over the complete phase space, then the procedure should be the
following. The phase space should be divided into regions where one of the
two calculations is applicable. The calculation then is to be performed in
the respective regions, assuming extreme scale hierarchy everywhere in
that region.  Then the sum will not depend on the cutoff defining the
regions at the order at which calculation is performed.  It is not
possible to perform one calculation, which would have the same accuracy in
all the different regions of the phase space with a different scale
structure.  In what follows we will be interested only in contributions
leading in $\Lt/\Lo$, which corresponds to DPA.  The possibility of using
the effective field theory framework to go beyond DPA is interesting and
very relevant, but it is still an open problem.

\subsection{Integrating out resonant modes}
\label{sec:eff/resonant_neutral}

Starting from the effective Lagrangian~\Eqn{eff/hard}, let us now
integrate out resonant modes, $\phi(P_M,p)$, $\psi(P_0,p)$. These modes
all have momenta $P_M\sim P_0\sim\Lo$, $p\sim \Lt$, $D\sim\Lo\Lt$.  This
will result in another effective Lagrangian that contains soft modes,
$\psi(p)$ and $A_{\mu}(p)$, with momenta $p\sim\Lt$, $D\sim\Lt^2$, and external modes,
$\psi(P_0)$, with momenta $P_0\sim\Lo$, $D=0$. The importance of operators in
the effective Lagrangian is determined by powers of
$\Lt^2/\Lo\Lt=\Lt/\Lo$.  The operators in the new effective Lagrangian are
nonlocal.  The leading operator corresponding to pair production of
unstable particles is a $\psi^4$-operator and has the form
\be
\label{eff/resonant/born}
        \Lambda(x,y,z) \ \cdot \ 
        (\bar{\psi} \DD \psi)(x) \ \cdot \ 
        X\, \PP(y) \ \cdot \ 
        (\bar{\psi} \DD \psi)(z).
\ee
This term describes the resonant Born diagram in the leading
approximation.  The factor $\PP(y)$ corresponds to the production of
unstable particles at the point $y$ and $(\bar{\psi} \DD \psi)(x)$
corresponds to their decay at the point $x$.  As long as the unstable
particles are neutral, each of these factors is gauge invariant
separately.  The coefficient $\Lambda(x,y,z)$ contains derivatives with
respect to $x,y$ and $z$, which we will denote by $P_x$, $P_y$, and $P_z$
respectively. It describes the propagation of unstable particles from the
production to the decay point and has dimension $[-4]$. Only the
virtuality of modes that have been integrated out can appear in the
denominator. Thus, $\Lambda$ can be estimated as
\be
\label{eff/lambda_est}
        \Lambda(x,y,z)\sim\frac{1}{(\Lo\Lt)^2} 
        \Biggl(1+\OO\bigg(\frac{\Lt}{\Lo}\biggr)\Biggr).
\ee
As in section~\ref{sec:eff/hard} we will neglect $\Lt/\Lo$ suppressed
terms. Of course in this case we know the precise form of $\Lambda$. From
\Eqn{eff/hard} we deduce that the propagator of a $\phi(P_M,p)$-field is
given by $(2 P_M p + i M \Gamma)^{-1}$. This entails
\be
\Lambda(x,y,z) = \frac{1}{2 P_x p_x+iM\Gamma} \ \cdot \
                \frac{1}{2 P_z p_z+iM\Gamma}.
\ee 
When the operator $P_x$ acts on a fermion field, $\psi(P)(x)$, it gives
its momentum $P$, when it acts on pair of fermion fields,
$(\bar{\psi}(P_1) \psi(P_2))(x)$, it gives the sum of the momenta,
$P_1+P_2$.  The total momentum of the fermion pair is equal to the
momentum of the unstable particle.  Therefore it has hard and soft
components, $(P_1+P_2)=P+p$.  Because we integrated out the unstable
particle field the effective Lagrangian does not depend on the unstable
particle field itself any longer.  Nevertheless, it can depend of its
momentum. 

Of course, in more complicated cases the coefficients in front of the
nonlocal operators can be quite complicated. An explicit matching
calculation is required in order to determine them.
In Sect.~\ref{sec:eff/resonant_neutral/matching/QED} and
Sect.~\ref{sec:eff/resonant_neutral/matching/QCD} we write down the nonlocal operators
explicitly and show how the nonlocal Wilson coefficients can be determined by the  matching.
Our main goal, however, is to reproduce the estimates of non-factorizable
corrections presented in Section~\ref{sec:feyn}. 
In order to do that we will not
need to know the explicit form of the nonlocal coefficients. The estimates
of various quantities can be determined by the scaling properties alone.
In Sect.~\ref{sec:eff/resonant_neutral/estimates} we show how the scaling properties
lead to such estimates.

\subsubsection{Explicit operators and matching in QED}
\label{sec:eff/resonant_neutral/matching/QED}

We start by writing down an effective Lagrangian describing pair
production of unstable particles. The Lagrangian will be written in
terms of (nonlocal) gauge invariant operators dependent on soft
fields, $\psi(p)$ and $A_{\mu}(p)$, and external fields $\psi(P_0)$.
The soft fields transform locally under gauge transformations whereas
the external fields transform globally.  We will perform the matching
by comparing matrix elements squared calculated in the effective
theory and in the underlying theory.%
\footnote{Matching matrix 
   elements would be incorrect because this does not fully reflect the
   fact that $\psi(P_0)$ are external fields by construction.  In
   particular, the color traces have to be taken, because $\psi(P_0)$
   is subject to global gauge transformation only.}

The effective Lagrangian consists of kinetic terms and terms
describing interactions.  It has the following structure
\begin{eqnarray}
	\LL_{\sss{eff}}
&=&
	- \frac{1}{4} F_{\mu \nu} F^{\mu \nu}
	+ \bar{\psi}(p) (\not\!P_0\,+\!\not\!\cov{p}\,)  \psi(p) 
	+ \bar{\psi}(P_0)\!\not\!P_0\, \psi(P_0)
\nonumber \\
\label{eff/resonant}
&&
        +\
	\Lambda(x,y,z) \ \cdot \ 
        ( \bar{\psi}\DD_{\sss{int}} \psi)(x) \ \cdot \ 
        X\, \PP_{\sss{int}}(y) \ \cdot \ 
        (\bar{\psi} \DD_{\sss{int}} \psi)(z),
\end{eqnarray}
where the first three terms are kinetic terms. Note that $\cov{p}\,$
contains the covariant derivative whereas $p$ contains the normal
derivative. Therefore, the soft fermions $\psi(p)$ do interact with
the soft gauge field $A_{\mu}(p)$ but there is no interaction of the
external field, $\psi(P_0)$, with $A_{\mu}(p)$.  The last term in
\Eqn{eff/resonant} describes doubly resonant production-decay
processes.  The field $\psi$ appearing there is an external field
$\psi(P_0)$.  Of course, one can easily generalize this Lagrangian to
describe production of more than two unstable particles.  The
operators $\DD_{\sss{int}}$ and $\PP_{\sss{int}}$ contain operators
with different powers of the gauge field $A_\mu$, describing gauge
boson radiation accompanying the production/decay process.
Schematically, they have the following structure
\begin{eqnarray}
  \PP_{\sss{int}} &=& 
  \PP\biggl(1 + e F_{\mu\nu} + e^2 (F_{\mu\nu})^2 + \ldots\biggr) 
  \label{eff/fullP}  \\
  \DD_{\sss{int}} &=&
  \DD\biggl(1 + e F_{\mu\nu} + e^2 (F_{\mu\nu})^2 + \ldots\biggr).
  \label{eff/fullD} 
\end{eqnarray}
The first terms correspond to production/decay without radiation. The
terms linear in $A_\mu$ correspond to production/decay accompanied by
radiation of one gauge boson, and so on.  If the source $X$ and the
unstable particles is neutral then there there can be no radiation
from the production stage of the process
\be
   \PP_{\sss{int}}=\PP.
   \label{eff/simpleP}
\ee

Let us start with the QED case.  We will write down explicitly
operators corresponding to different terms in
$(\bar{\psi}\DD_{\sss{int}} \psi)(x)$.  The operator corresponding to
one photon radiation is
\be
\label{eff/operator_F/QED}
	(\bar{\psi}\DD_{\sss{int}} \psi)(X)\biggl.\biggr|_{\sim F} 
	=
	\int dx dy dz \ 
	\Sigma^{(1)}_{\mu\nu}(x-X,y-X|z-X) \ 
	\Bigl(\bar{\psi}(x) \DD \psi(y)\Bigr) \ 
	F^{\mu\nu}(z).
\ee
This operator is gauge invariant because $\psi$ is an external field
subject to a global gauge transformation and $A_{\mu}$ is an abelian
gauge field.  This operator contributes only to one photon radiation.
It is convenient to use a Fourier transformed function for the Wilson
coefficient 
\be
	\Sigma^{(1)}_{\mu\nu}(x-X,y-X|z-X)
	=
	\int d^3 s \ 
	e^{is_1(x-X)}e^{is_2(y-X)}e^{is_3(z-X)} \ 
	\Sigma^{(1)}_{\mu\nu}(s_1,s_2|s_3).
\ee
The matching calculation in the effective theory and in QED is
\be
 \begin{picture}(40,30)(0,-2) 
	\GCirc(0,0){3}{0}
	\put(-15,-5){\small $X$}
	\Photon(0,0)(0,25){2}{5}
	\put(-20,20){\small $l,\alpha$}
	\ArrowArc(20,-20)(28.3,40,140)
	\put(10,10){\small $k$}
	\ArrowArc(20,20)(28.3,-140,-40)
	\put(10,-20){\small $k'$}
	\GCirc(40,0){3}{0}
        \DashLine(20,-25)(20,25){3}
\end{picture}
\rule[-30pt]{0mm}{0mm}
	\ \sim \
	e^{iX(k+k'+l)}
	\Sigma^{(1)}_{\mu\nu}(k,k'|l) \ 
	(l^{\mu}g^{\nu\alpha}-l^{\nu}g^{\mu\alpha})
	=
	e \ e^{iX(k+k'+l)}
	\biggl(\frac{k}{kl}-\frac{k'}{k'l}\biggr)^{\alpha}.
\label{QEDmatch}
\ee
Here we recognize the non-factorizable current structure that we
encountered in the QED calculations of Sect.~\ref{sec:feyn}. The Wilson
coefficient can be read off from \Eqn{QEDmatch} and is given by
\be
	\Sigma^{(1)}_{\mu\nu}(k,k'|l) = \frac{e}{2} \ 
\left(\frac{k_\mu k^{\prime}_{\nu}}{kl\ k'l}
- \frac{k^{\prime}_{\mu} k_\nu }{kl\ k'l} \right)
\ee
This result generalizes to $N$-photon radiation in a straightforward
way.  The operator responsible for radiation of $N$ photons is
\bea
	(\bar{\psi}\DD_{\sss{int}} \psi)(X)\biggl.\biggr|_{\sim F^N} 
	&=&
	\int dx dy d^Nz \ 
	\Sigma^{(N)}_{\mu_1\nu_1\ldots\mu_N\nu_N}(x-X,y-X|z_1-X,\ldots,z_N-X)
\label{eff/operator_NF/QED}
\\ \nonumber
 && \qquad \qquad \times \
	\Bigl(\bar{\psi}(x) \DD \psi(y)\Bigr) \ 
	F^{\mu_1\nu_1}(z_1) \ \ldots \ F^{\mu_N\nu_N}(z_N).
\eea
This operator is gauge invariant and contributes solely to observable
where the number of emitted photons is $N$. The matching calculation
in QED is very similar to the calculations presented in
Sect.~\ref{sec:feyn} with a product of $N$ non-factorizable currents
emerging as a result.  The resulting Wilson coefficient is given by
\be
\label{wilson_abelian}
	\Sigma^{(N)}_{\mu_1\nu_1\ldots\mu_N\nu_N}(k,k'|l_1\ldots l_N) = 
	\frac{e^N}{2^N N!} \ 
   \left( \frac{k_{\mu_1} k^{\prime}_{\nu_1}}{kl_1\ k'l_1} -
          \frac{k^{\prime}_{\mu_1} k_{\nu_1}}{kl_1\ k'l_1} \right)
	\ldots	  
  \left( \frac{k_{\mu_N} k^{\prime}_{\nu_N}}{kl_N\ k'l_N} -
         \frac{k^{\prime}_{\mu_N} k_{\nu_N}}{kl_N\ k'l_N} \right)\ .
\ee

We can summarize the situation as follows: the propagation of soft and
external fields is described by the kinetic operators in
\Eqn{eff/resonant}. The production and decay of a pair of unstable
particles accompanied by radiation of an arbitrary number of photons
is described by the last term in \Eqn{eff/resonant} with the factors
$(\bar{\psi}\DD_{\sss{int}} \psi)(x)$ explicitly given in
\Eqn{eff/operator_NF/QED}. These operators reproduce those non-factorizable
corrections that are due to interaction with decay dipoles and have
the non-factorizable current structure encountered in
Sect.~\ref{sec:feyn}. As mentioned in Sect.~\ref{sec:feyn} there are
also non-factorizable corrections due to the coupling of soft fermions
to soft gauge bosons, which induces corrections due to
propagation. Since soft fermions as well as soft gauge boson fields
are still dynamical in the Lagrangian of the effective theory,
\Eqn{eff/resonant}, these corrections are absolutely identical in the
effective and underlying theory.

\subsubsection{Explicit operators and matching in QCD}
\label{sec:eff/resonant_neutral/matching/QCD}

We now want to generalize the construction of the effective theory to
the non-abelian case of QCD.  The effective Lagrangian has a structure
similar to the one encountered in the QED case, \Eqn{eff/resonant}.
However, the explicit form of the operators
$(\bar{\psi}\DD_{\sss{int}} \psi)(x)$ has to change now.  Indeed,
the operator responsible for one photon radiation,
\Eqn{eff/operator_F/QED}, is not gauge invariant in the non-abelian
case. This is because the field tensor $F_{\mu\nu}$ is not gauge
invariant if $A_\mu$ is a non-abelian gauge field. In fact it is
impossible to construct a gauge invariant operator that contains
$\bar{\psi}\psi A_\mu$ if $\psi$ is transforms globally and $A_\mu$
transforms locally under gauge transformations.  This is
consistent with what is observed in QCD. It merely reflects the fact
that the emission of a single gluon is forbidden in QCD due to the
vanishing of the color trace.   The matching equation for the
emission of a single gluon then is simply
\be
 \begin{picture}(40,30)(0,-2) 
	\GCirc(0,0){3}{0}
	\put(-15,-5){\small $X$}
	\Photon(0,0)(0,25){2}{5}
	\put(-30,20){\small $l,\alpha,a$}
	\ArrowArc(20,-20)(28.3,40,140)
	\put(10,10){\small $k$}
	\ArrowArc(20,20)(28.3,-140,-40)
	\put(10,-20){\small $k'$}
	\GCirc(40,0){3}{0}
        \DashLine(20,-25)(20,25){3}
\end{picture}
\rule[-30pt]{0mm}{0mm}
	\ = \ 0 \, .
\label{eff/operator_1F/QCD}
\ee

The first non-vanishing process in QCD is the one where two gluons are
radiated. Accordingly, the first non-trivial gauge invariant operator
in the effective theory contains two $F_{\mu\nu}$ tensors
\bea
	(\bar{\psi}\DD_{\sss{int}} \psi)(X)\biggl.\biggr|_{\sim F^2} 
	&=&
	\int dx dy d^2z \ 
	\Sigma^{(2)}_{\mu\nu\sigma\rho}(x-X,y-X|z_1-X,z_2-X) \ 
\label{eff/operator_2F/QCD}
\\ \nonumber &\times&
	\Bigl(\bar{\psi}(x) \DD \psi(y)\Bigr) \ 
	\Tr \biggl(F^{\mu\nu}(z_1) U(z_1,z_2) F^{\sigma\rho}(z_2)
	U(z_2,z_1)\Biggr). 
\eea
Here we introduced $U(z_1,z_2)$, the path-ordered exponential defined as
\be
	 U(x,y) = 
	\mbox{P}\exp\Biggl[ 
             -\,ig\int\limits_x^y A_{\mu}(\omega)\,d\omega^{\mu}
	\Biggr] \, ,
\ee
in order to ensure the gauge invariance of the operator.  The matching
calculation in the effective theory and in  QCD for the two gluon
process is similar to the one in QED
\begin{eqnarray}
 \begin{picture}(40,30)(0,-2) 
	\GCirc(0,0){3}{0}
	\put(-15,-5){\small $X$}
	\Photon(0,0)(-5,25){2}{5}
	\Photon(0,0)(+5,25){2}{5}
	\ArrowArc(20,-20)(28.3,40,140)
	\put(10,10){\small $k$}
	\ArrowArc(20,20)(28.3,-140,-40)
	\put(10,-20){\small $k'$}
	\GCirc(40,0){3}{0}
        \DashLine(20,-25)(20,25){3}
\end{picture}
\rule[-30pt]{0mm}{0mm}
	\ 
&\sim&  \
	\Tr\bigl(T^{a}T^{b}\Bigr) \ 
	\Sigma^{(2)}_{\mu\nu\sigma\rho}(k,k'|l_1,l_2) \ 
	(l_1^{\mu}g^{\nu\alpha}-l_1^{\nu}g^{\mu\alpha})
	(l_2^{\sigma}g^{\rho\beta}-l_2^{\rho}g^{\sigma\beta})
	+ (1\leftrightarrow 2)
\nonumber \\
	\ 
&=& \ 
	g^2 \ 
	\Tr\bigl(T^{a}T^{b}\Bigr) \ 
	\biggl(\frac{k}{kl_1}-\frac{k'}{k'l_1}\biggr)^{\alpha} \ 
	\biggl(\frac{k}{kl_2}-\frac{k'}{k'l_2}\biggr)^{\beta}.
\label{QCDmatch2}
\end{eqnarray}
The solution for the Wilson coefficient can be read off from
\Eqn{QCDmatch2} 
\be
	\Sigma^{(2)}_{\mu\nu\sigma\rho}(k,k'|l_1,l_2) 
	= \frac{g^2}{2^2 2!} \
     \left( \frac{k_\mu k^{\prime}_{\nu}}{kl_1\ k'l_1} -
            \frac{k^{\prime}_{\mu} k_\nu}{kl_1\ k'l_1} \right)\
     \left( \frac{k_\sigma k^{\prime}_{\rho}}{kl_2\ k'l_2} -
            \frac{k^{\prime}_{\sigma}k_\rho}{kl_2\ k'l_2} \right)\ .
\ee
This is similar to what we found in QED. However, there is an
important difference compared to the QED case in that now the operator
that is responsible for the emission of two gluons,
\Eqn{eff/operator_2F/QCD}, will contribute also to processes with a
larger number of gluons radiated off. To illustrate this point, let us
consider a process with the emission of three gluons. The diagrams of
the effective theory contributing to this process are shown in
Fig.~\ref{fig:eff/3gluon_diagrams}.  There will be three distinct
contributions.
\begin{figure}[ht]
\bigskip
\bigskip
\begin{center}
 \begin{picture}(40,30)(0,-30) 
    	\GCirc(0,0){3}{0}
	\put(-25,-5){\small $\Sigma^{(2)}$}
	\Photon(0,0)(-10,25){2}{5}
	\Photon(0,0)(+5,15){2}{3}
	\Photon(5,15)(10,25){2}{3}
	\Photon(5,15)(0,25){2}{3}
	\ArrowArc(20,-20)(28.3,40,140)
	\ArrowArc(20,20)(28.3,-140,-40)
	\GCirc(40,0){3}{0}
        \DashLine(20,-25)(20,25){3}
	\put(0,-30){(a)}
\end{picture} 
\phantom{XXXXXXX}
 \begin{picture}(40,60)(0,-30) 
    	\GCirc(0,0){3}{0}
	\put(-25,-5){\small $\Sigma^{(2)}$}
	\Photon(0,0)(-10,25){2}{5}
	\Photon(0,0)(+10,25){2}{5}
	\Photon(0,0)(+0,25){2}{5}
	\ArrowArc(20,-20)(28.3,40,140)
	\ArrowArc(20,20)(28.3,-140,-40)
	\GCirc(40,0){3}{0}
        \DashLine(20,-25)(20,25){3}
	\put(0,-30){(b)}
\end{picture} 
\phantom{XXXXXXX}
 \begin{picture}(40,60)(0,-30) 
    	\GCirc(0,0){3}{0}
	\put(-25,-5){\small $\Sigma^{(3)}$}
	\Photon(0,0)(-10,25){2}{5}
	\Photon(0,0)(+10,25){2}{5}
	\Photon(0,0)(+0,25){2}{5}
	\ArrowArc(20,-20)(28.3,40,140)
	\ArrowArc(20,20)(28.3,-140,-40)
	\GCirc(40,0){3}{0}
        \DashLine(20,-25)(20,25){3}
	\put(0,-30){(c)}
\end{picture} 
\ccaption{}{Diagrams of the effective theory contributing to the three
    	gluon process. \label{fig:eff/3gluon_diagrams}}  
   \end{center}
\end{figure}
In the first one, shown in Fig.~\ref{fig:eff/3gluon_diagrams}(a), two
gluons are emitted due to the $(\bar{\psi}\psi)F^2$ operator,
\Eqn{eff/operator_2F/QCD}, with one of the gluons subsequently
splitting into two due to three-gluon interaction coming from the
kinetic term of the gauge field, $-1/4\ F_{\mu \nu}F^{\mu \nu}$. The
second is shown in Fig.~\ref{fig:eff/3gluon_diagrams}(b) and is also
due to the $(\bar{\psi}\psi)F^2$ operator. There are two sources in
$(\bar{\psi}\psi)F^2$ that give rise to terms contributing to three
gluon radiation. First, the non-abelian commutator terms in
$F_{\mu\nu}$ result in terms with three (and four) gauge
fields. Second, the expansion of the path-ordered exponentials
$U(z_1,z_2)$ creates terms with an arbitrarily large number of $A_\mu$
fields. Thus, the operator $(\bar{\psi}\psi)F^2$ given in
\Eqn{eff/operator_2F/QCD} contributes to all processes with two or
more gluons emitted. Finally, the third contribution to the three
gluon process is shown schematically in
Fig.~\ref{fig:eff/3gluon_diagrams}(c).  This contribution is due to a
new $\bar{\psi}\psi F^3$ operator with a Wilson coefficient
$\Sigma^{(3)}$ defined as
\bea
	(\bar{\psi}\DD_{\sss{int}} \psi)(X)\biggl.\biggr|_{\sim F^3} 
	&=&
 \int dx dy d^3z \ 
 \Sigma^{(3)}_{\mu_1\nu_1\mu_2\nu_2\mu_3\nu_3}(x-X,y-X|z_1-X,z_2-X,z_3-X) 
\label{eff/operator_3F/QCD}
\\ \nonumber
 && \hspace*{-1.0cm} \times \
	\Bigl(\bar{\psi}(x) \DD \psi(y)\Bigr) \ 
	\Tr \biggl(F^{\mu_1\nu_1}(z_1) U(z_1,z_2)
			F^{\mu_2\nu_2}(z_2) U(z_2,z_3)
			F^{\mu_3\nu_3}(z_3) U(z_3,z_1) \Biggr).
\eea
The sum of the three contributions shown in
Fig.~\ref{fig:eff/3gluon_diagrams} should be equivalent to the full
QCD calculation. This matching condition determines the Wilson
coefficient $\Sigma^{(3)}$.

This picture is the non-abelian version of what we found in QED.
Non-factorizable corrections are either due to the interaction of soft
gluons with the decay dipoles or due to propagation corrections of the
soft gluons. In the abelian case the propagation corrections arise
solely from interaction with soft fermions. In the non-abelian case
there is an additional source of propagation corrections, namely the
self-couplings of gluons.  What we achieved is a generalization
to the non-abelian case of a gauge invariant separation between
non-factorizable corrections due to interaction with decay dipole and
non-factorizable corrections due to propagation.  The analogy of the
estimates of Sect.~\ref{sec:feyn} for the non-factorizable corrections
due to the interaction with decay dipoles is now simply an estimate
for the Wilson coefficients $\Sigma^{(N)}$.  For example, the
contribution shown in Fig.~\ref{fig:eff/3gluon_diagrams}(c) is due to
a $\OO(g^3)$ dipole interaction. On the other hand, the contributions
shown in Fig.~\ref{fig:eff/3gluon_diagrams}(a) and (b) are due to
$\OO(g^2)$ dipole interactions plus $\OO(g)$ propagation corrections.
Thus, it is only within the effective theory framework that the true
meaning of the estimates for non-factorizable corrections in the
non-abelian case becomes clear.

The matching calculation in QCD is far more complicated than in QED.
What will be important for us here is that the Wilson coefficient 
can be estimated as
\be
\label{wilson_estimate}
	\Sigma^{(3)}_{\mu_1\nu_1\mu_2\nu_2\mu_3\nu_3}(k,k'|l_1,l_2,l_3) 
	\sim g^3  \frac{1} {(\Lo\Lt)^6}\ 
      k_{\mu_1}k'_{\nu_1}k_{\mu_2}k'_{\nu_2}k_{\mu_3}k'_{\nu_3}\ .
\ee

In the following section we concentrate on the estimates of the
non-factorizable corrections due to interactions with the decay
dipoles. It will be sufficient to know the estimates of the Wilson
coefficients, \Eqn{wilson_estimate}, as well as the Lorentz structure
and the form of the operators. The explicit form of the matching
coefficients will not be important. Nevertheless, we performed an
explicit matching calculation for $\Sigma^{(3)}$ in order to convince
the reader that the effective theory does exist and can be matched to
QCD. We should mention that the set of operators we have chosen is
sufficient only at leading order in $\Lt/\Lo$. If we were to go beyond
this approximation additional operators with a more complicated
$\gamma$-matrix structure would be needed. Details of the matching
calculation are given in Appendix~\ref{app:b}.

\subsubsection{Scaling properties and estimates}
\label{sec:eff/resonant_neutral/estimates}

In this subsection we will generalize the estimates of the Wilson
coefficients found above and then apply these in order to obtain
estimates of the non-factorizable correction. Thereby we will confirm
the results that have been found in Section~\ref{sec:feyn} by
analyzing Feynman diagrams.

Instead of working with the momenta of the fermions it will turn out to be
more convenient to work with the momenta $P_M+p$ and $\Delta$, defined as
follows:
\be
  P_M+p=P_1+P_2, \ \ \ \Delta \equiv k-k', \ \ \ P_M\Delta=-p\Delta,
  \label{eff/Deltadef}
\ee
where $P_M$ is the hard on-shell momentum of the resonant $\phi$-field
that decays into the fermions with momentum $k$ and $k'$. $p$ is
the corresponding soft momentum, $p\sim\Lt$. Finally, $\Delta$ is a
hard off-shell momentum, $\Delta\sim\Lo$, however $P_M\Delta\sim\Lo\Lt$.

The estimate found in \Eqn{wilson_estimate} can be generalized to the
Wilson coefficient $\Sigma^{(n)}$ of the operator $(\bar{\psi}\psi)F^n$
\be
\Sigma^{(n)}_{\mu_1\nu_1\ldots\mu_n\nu_n} \sim g^n 
      \frac{1} {(\Lo\Lt)^{2n}}\ 
      k_{\mu_1}k'_{\nu_1} \ldots k_{\mu_3}k'_{\nu_3}\ 
      \sim \frac{g^n}{\Lt^{2n}}\ .
\label{allwilson_estimate}
\ee
Rewriting this in terms of the new momenta we find 
\be
\label{eff/selection_rule}
 (\bar{\psi}\DD_{\sss{int}} \psi)(P_M,p,\Delta)\biggl.\biggr|_{\sim F^n} 
  \sim
  \frac{(P_M \ \mbox{or} \ \Delta)^{2n+m} p^{m}}{(\Lo\Lt)^{2n+m}}\ 
  (\bar{\psi}\DD\psi)\ (eF)^n 
\ee
We would like to stress again here that this selection rule is valid only
up to $\Lt/\Lo$ suppressed terms.  

The selection rule, \Eqn{eff/selection_rule}, can be understood as
follows: the amplitude for radiating one soft gauge boson is
proportional to $g P_M^\mu/(P_Mp)\sim g/\Lt$, plus $\Lt/\Lo$ suppressed
contributions.  Consequently, the Feynman rules corresponding to such
an interaction operator should scale as $\sim g/\Lt$.  This means that
the term linear in $gA_{\mu}$ should scale as $gA/\Lt$.  As $A_\mu$ is
a soft field, this is equivalent to terms linear in $gF_{\mu\nu}$
scaling as $gF/\Lt^2$.  These arguments can be extended to radiation
of $n$-gauge bosons. This leads to the observation that $(gF)^n$-terms
have to scale as $(gF)^n/\Lt^{2n}$, as stated in
\Eqn{allwilson_estimate}. Taking into account that only $\Lo\Lt$
can appear in the denominator we balance the dimension of the
contribution of the operators by powers of $\Lo$ and $\Lt$ in the
numerator. Powers of $\Lo$ correspond to hard momenta $P$ or $\Delta$
whereas powers of $\Lt$ correspond to soft momenta $p$.  This leads
directly to the selection rule \Eqn{eff/selection_rule}.

Let us use this selection rule to study the non-factorizable
correction due to the exchange of one soft $A_\mu$ field. As mentioned
above, \Eqn{eff/operator_1F/QCD}, this contribution will vanish in QCD
and our estimate will only be relevant for QED. The Feynman diagram
corresponding to this one-loop correction is shown in
Fig.~\ref{fig:eff/resonant/corr}(a). According to
\Eqn{eff/selection_rule}, the leading contribution to the Feynman rule
for the vertex scales as 
\be
\begin{picture}(30,20)(0,-2) 
        \GCirc(0,0){3}{0}
        \ArrowLine(0,0)(30,10)
        \ArrowLine(30,-10)(0,0)
        \Photon(0,0)(15,15){2}{4}
        \DashCArc(0,0)(35,-20,20){1}
\end{picture}
        \ \ \ 
        (P_M,p,\Delta)
        \ \sim 
         \DD \ e\frac{1}{\Lt}.
\label{vertex1est}
\ee
More explicitly, the following operators linear in $F_{\mu\nu}$ are
possible
\be
\label{eff/excl/1}
        (\bar{\psi}\DD_{\sss{int}} \psi)(P_M,p,\Delta)\biggl.\biggr|_{\sim F} =
        \biggl[ 
        \frac{P^\mu_{M}\Delta^\nu}{\Lo\Lt}
        +
        \frac{(\Delta^{\mu} p^\nu+P^\mu_{M} p^\nu)}{\Lo\Lt}
        \frac{P_M^2+\Delta^2}{\Lo\Lt}
        +\ldots
        \biggr] 
        \ \cdot \ 
        \frac{e}{\Lo\Lt}
        (\bar{\psi}\DD \psi) F_{\mu\nu} ,
\ee
where in the last term we used the fact that $P_M\Delta$ is
subleading. The leading contribution to the Feynman rule is coming
from the first terms in \Eqn{eff/excl/1}. With the help of
\Eqn{vertex1est} the one-loop correction relative to the Born term can
be estimated as
\be
        \delta_1 \sim
        \int \frac{d^4l}{l^2} \ \frac{e}{\Lt} \  \frac{e}{\Lt}
        \sim
        \frac{\Lt^4}{\Lt^2} \ \frac{e}{\Lt} \ \frac{e}{\Lt}
        \sim e^2 ,
\ee
where $l$ is the soft momentum of the exchanged gauge boson. This
shows that non-factorizable corrections to exclusive quantities are
$\OO(\alpha)$ and are not suppressed by any additional parameters.
\begin{figure}[ht]
\bigskip
\bigskip
\begin{center}
\begin{picture}(80,60)(0,-30) 
        \GCirc(0,0){3}{0}\put(5,5){$y$}
        \GCirc(30,20){3}{0}\put(17,20){$x$}
        \ArrowLine(30,20)(60,30)
        \ArrowLine(60,10)(30,20)
        \GCirc(30,-20){3}{0}\put(17,-20){$z$}
        \ArrowLine(30,-20)(60,-10)
        \ArrowLine(60,-30)(30,-20)
        \Photon(30,20)(30,-20){2}{6}
        \put(-10,-30){(a)}
\end{picture}
\phantom{XXXXXXXXXX}\begin{picture}(80,60)(0,-30) 
        \GCirc(0,0){3}{0}\put(5,5){$y$}
        \GCirc(30,20){3}{0}\put(17,20){$x$}
        \ArrowLine(30,20)(60,30)
        \ArrowLine(60,10)(30,20)
        \GCirc(30,-20){3}{0}\put(17,-20){$z$}
        \ArrowLine(30,-20)(60,-10)
        \ArrowLine(60,-30)(30,-20)
	\PhotonArc(60,0)(36.05,150,210){-1.5}{6}	
	\PhotonArc(0,0)(36.05,-30,30){1.5}{6}	
        \put(-10,-30){(b)}
\end{picture}
\ccaption{}{Feynman diagrams for one-loop (a) and two-loop (b)
  non-factorizable corrections in an effective theory where resonant
  modes have been integrated out.  \label{fig:eff/resonant/corr}}
  \end{center}
\end{figure}

In order to compare to our previous results, let us recall that the
estimates of non-factorizable corrections presented in
Sect.~\ref{sec:feyn} are valid for a special class of observables
only. The estimates are valid if one integrates out part of the phase
space corresponding to final the state fermions, keeping the momenta
and invariant masses of the unstable particles fixed.  This
corresponds to the construction of operators in the effective field
theory using $P_M^\mu$ and $p^\mu$, but not $\Delta^\mu$.  Indeed,
suppose that the contribution of an operator depends on
$\Delta$. After performing a phase-space integral over the fermion
decay angles this contribution will be expressed in terms of $P_M$ and
$p$ only. Thus, in order to reproduce the estimates of
Sect.~\ref{sec:feyn} we will have to construct a new effective
Lagrangian. This Lagrangian is the same as the previous one, but
without terms containing $\Delta$. This Lagrangian will be used to
calculate observables, in which the integration over fermion decay
angles is already performed. Consulting \Eqn{eff/excl/1}, the leading
operator linear in $F_{\mu\nu}$ is 
\be
\label{eff/incl/1}
        (\bar{\psi}\DD_{\sss{int}} \psi)(P_M,p)\biggl.\biggr|_{\sim F} =
        e\
        \frac{P^\mu_{M} p^\nu P_M^2}{(\Lo\Lt)^3} \ 
        (\bar{\psi}\DD\psi) \ F_{\mu\nu}.
\ee
Therefore, the Feynman rule corresponding to the
one-gauge-boson-radiation vertex  in this new Lagrangian scales as 
$$
\begin{picture}(30,20)(0,-2) 
        \GCirc(0,0){3}{0}
        \ArrowLine(0,0)(30,10)
        \ArrowLine(30,-10)(0,0)
        \Photon(0,0)(15,15){2}{4}
        \DashCArc(0,0)(35,-20,20){1}
\end{picture}
        \ \ \ 
        (P_M,p)
        \ \sim 
        \DD \ e\frac{1}{\Lt}\frac{P_M^2}{\Lo^2}.
$$
The estimate of the non-factorizable correction in this case is 
\be
\label{eff/est/1}
        \delta_1 \sim
        \frac{\Lt^4}{\Lt^2} \ \cdot \ 
        \frac{e}{\Lt}\frac{P_M^2}{\Lo^2} \ \cdot \ 
        \frac{e}{\Lt}\frac{P_M^2}{\Lo^2}
        \sim e^2 \frac{M^4}{E^4}.
\ee
This is an estimate similar to what we had in Sect.~\ref{sec:feyn}.  It
shows that at high energy, $E\gg M$, non-factorizable corrections are
suppressed.

Of course, it can be that the underlying theory has more
symmetries. In that case one has to take them into account in the
construction of the relevant operators. This is the reason why the
estimate obtained in \Eqn{eff/est/1} does not agree with
\Eqns{feyn:nfneutraleven}{feyn:nfneutralodd}. Indeed, if the unstable
particle is neutral and it decays into a fermion-antifermion pair with
the same flavour there is an antisymmetry under the exchange of final
state particles.  This leads to the statement that the one-loop
non-factorizable correction is zero in this case.  In fact, all
odd-order-loop non-factorizable corrections are zero because of this
symmetry.  In the effective theory language this means that operators
linear in $F$, like the one given in \Eqn{eff/incl/1} are
forbiden. Alternatively, performing the matching one will find that
the corresponding Wilson coefficients vanish. However, if there is no
additional symmetry the estimate given in \Eqn{eff/est/1} holds.

Let us extend this analysis to two-loop corrections. The explicit form
of the terms quadratic in $F_{\mu\nu}$ is
\be
\label{eff/incl/2}
        (\bar{\psi}\DD_{\sss{int}} \psi)(P_M,p)\biggl.\biggr|_{\sim F^2}=
        e^2\ 
        \frac{1}{(\Lo\Lt)^4}
        \biggl(g^{\nu\sigma} P_M^{\mu} P_M^{\rho} P_M^2
                +
                g^{\nu\sigma}g^{\mu\rho} P_M^4 \biggr)
        (\bar{\psi} \ \DD\psi) \ F_{\mu\nu}F_{\sigma\rho} .
\ee
The diagram corresponding to two-loop correction is shown in
Fig.\ref{fig:eff/resonant/corr}(b). The Feynman rule for the vertex
with the emission of two gauge bosons scales like
\be
\begin{picture}(30,20)(0,-2) 
        \GCirc(0,0){3}{0}
        \ArrowLine(0,0)(30,10)
        \ArrowLine(30,-10)(0,0)
        \Photon(0,0)(10,15){2}{3}
        \Photon(0,0)(24,15){2}{5}
        \DashCArc(0,0)(35,-20,20){1}
\end{picture}
        \ \ \ 
        (P_M,p)
        \ \sim \ \ 
        \DD \ 
        \frac{e^2}{\Lt^2}\cdot \frac{P_M^2}{\Lo^2}
        \biggl(1+ \frac{P_M^2}{\Lo^2}\biggr).
\ee
At high energy the leading contribution comes from the $g^{\nu\sigma}
P_M^{\mu} P_M^{\rho} P_M^2$ term and the contribution of the
$g^{\nu\sigma}g^{\mu\rho} P_M^4$ term is suppressed by $M^2/E^2$.  The
relative two-loop non-factorizable correction can be estimated as
\be
        \delta_2 \sim
        \int \frac{d^4k_1}{k_1^2} \frac{d^4k_2}{k_2^2} \ 
        \biggl(\frac{e^2}{\Lt^2}\frac{P_M^2}{\Lo^2}\biggr) \ 
        \biggl(\frac{e^2}{\Lt^2}\frac{P_M^2}{\Lo^2}\biggr)
        \sim
        \biggl(\frac{\Lt^4}{\Lt^2}\biggr)^2 \ 
        \biggl(\frac{e}{\Lt^2}\frac{P_M^2}{\Lo^2}\biggr) \ 
        \biggl(\frac{e}{\Lt^2}\frac{P_M^2}{\Lo^2}\biggr)
        \sim e^4 \  \frac{M^4}{E^4}.
\ee
This is precisely the estimate for two-loop non-factorizable final-final
corrections obtained in Sect.~\ref{sec:feyn}, \Eqn{feyn:est_even}.

It is possible to construct $N$th order terms in $F_{\mu\nu}$ for
$\DD_{\sss{int}}$.  It is clear that at high energy the leading
contribution will come from operators in which as many $P_M^\mu$ as
possible are contracted with $F_{\mu\nu}$.  This is because the
remaining $P_M^\mu$ needed for dimensional reasons will have to be
contracted with each other. Since $P_M^2\sim M^2\ll\Lo^2$, this
results in a suppression at high energy.  If $N$ is even
$(F_{\mu\nu})^N$ has $2N$ indices to be contracted. Due to the
antisymmetry of the field tensor a maximum of $N$ of these indices can
be contracted with $P_M^\mu$. The remaining $P_M^\mu$ have to be
contracted with each other.  By looking at the selection rule,
\Eqn{eff/selection_rule}, it can be seen that different $m$ are
possible.  All of them give the same contribution if $p_\mu$ is
contracted with $P_M^\mu$.  If $p^\mu$ is contracted with $F_{\mu\nu}$
then some extra $P_M^\mu$ would have to be contracted with each other
giving additional $M/E$ suppression.  This means that the leading
correction is the same as that coming from the selection rule with
$m=0$
\be
        (\bar{\psi}\DD_{\sss{int}} \psi)(P_M,p)\biggl.\biggr|_{\sim F^N} 
        \sim
        \frac{P_M^{2N}}{(\Lo\Lt)^{2N}}\ 
        (\bar{\psi}\DD\psi) (eF)^N ,
        \ \ \ \ \ \ 
        N\ \mbox{even}.
\ee
Note that this is in perfect agreement with the explicit result of
$\Sigma^{(3)}$ given in Appendix~\ref{app:b}. 
If $N$ is odd $(F_{\mu\nu})^N$ has $2N$ indices to be contracted, out of
which a maximum of $N$ can be contracted with $P_M^\mu$. The rest of the
indices cannot be contracted with each other because their number is odd.
Thus, at least one $p_\mu$ is necessary. Again, in the selection rule
\Eqn{eff/selection_rule} different values of $m$ are possible. However, 
the leading contribution comes from $m=1$. All other values of $m$
give the same result or suppressed contributions.  The selection rule
\Eqn{eff/selection_rule} gives for odd $N$
\be
      (\bar{\psi}\DD_{\sss{int}} \psi)(P_M,p)\biggl.\biggr|_{\sim F^N} 
        \sim
        \frac{P_M^{2N+1} p}{(\Lo\Lt)^{2N+1}} \
        (\bar{\psi}\DD \psi) (eF)^N ,
        \ \ \ \ \ \ 
        N\ \mbox{odd}.
\ee
Thus, the operator $\DD_{\sss{int}}$ is of the following form
\begin{eqnarray}
\lefteqn{
        (\bar{\psi}\DD_{\sss{int}} \psi)(P_M,p) = } &&
\label{eff/incl/N} \\
&&=\
        (\bar{\psi}\DD \psi) 
        +
        e\ 
        \frac{P^\mu_{M} p^\nu P_M^2}{(\Lo\Lt)^3} \ 
        (\bar{\psi}\DD\psi) \ F_{\mu\nu}
        +
        e^2\ 
        \frac{P^\mu_{M} P^\nu_{M} P_M^2}{(\Lo\Lt)^4}\ 
        (\bar{\psi}\DD\psi)  \ F_{\mu\sigma}F_{\nu}^{\ \sigma}
\nonumber        \\
&&+\     
        e^N\ 
        \frac{(P_M^2)^{N/2} P_M^{\mu_1}... P_M^{\mu_N}}{(\Lo\Lt)^{2N}} \ 
        T^{\sigma_1... \sigma_N} \ 
        (\bar{\psi}\DD\psi)\ F_{\mu_1\sigma_1}... F_{\mu_N\sigma_N}
        \biggl.\biggr|_{N=\sss{even}}
\nonumber \\
&&+\
        e^N\ 
        \frac{(P_M^2)^{(N+1)/2} P_M^{\mu_1}... P_M^{\mu_N} \ p^{\sigma_{N}}}
                                                {(\Lo\Lt)^{2N+1}} \ 
        T^{\sigma_1... \sigma_{(N-1)}} \ 
        (\bar{\psi}\DD\psi) \ F_{\mu_1\sigma_1}... F_{\mu_N\sigma_N}
        \biggl.\biggr|_{N=\sss{odd}}
        +
        \ldots,
\nonumber
\end{eqnarray}
where $T_{\sigma_1\ldots \sigma_N}$ is a symmetric tensor of even rank
constructed from $g_{\sigma_i\sigma_j}$.

We are now ready to estimate the $N$-loop non-factorizable correction.
The estimate will be different for odd and even $N$.  Let us start with
even $N$.  The Feynman rule corresponding to the radiation of $N$ gauge
bosons from the decay dipole is
\be
N\ \ \ 
\begin{picture}(60,20)(0,-2) 
        \GCirc(30,0){3}{0}
        \ArrowLine(30,0)(60,10)
        \ArrowLine(60,-10)(30,0)
        \Photon(30,0)(2,10){2}{6}
        \Photon(30,0)(2,-10){2}{6}
        \Photon(30,0)(0,4){2}{6}
        \Photon(30,0)(0,-4){2}{6}
        \DashCArc(30,0)(35,-20,20){1}
        \DashCArc(30,0)(35,160,200){1}
\end{picture}
        \ \ \ 
        (P_M,p)
        \ \sim \ \ 
        \DD \ 
        \frac{e^N}{\Lt^N}\frac{(P_M^2)^{N/2}}{\Lo^N},
        \ \ \ \ \ \ 
        N\ \mbox{even}.
\ee
Correspondingly, the relative $N$-loop non-factorizable correction is
\begin{eqnarray}
        \delta_N
&\sim&
        \int \frac{d^4k_1}{k_1^2}\ldots \frac{d^4k_N}{k_N^2} \ 
        \biggl(\frac{e^N}{\Lt^N}\frac{(P_M^2)^{N/2}}{\Lo^N}\biggr) \ 
        \biggl(\frac{e^N}{\Lt^N}\frac{(P_M^2)^{N/2}}{\Lo^N}\biggr)
\nonumber \\
&\sim&
        \biggl(\frac{\Lt^4}{\Lt^2}\biggr)^N \ 
        \biggl(\frac{e^N}{\Lt^N}\frac{(P_M^2)^{N/2}}{\Lo^N}\biggr)^2 \ 
        \sim e^{2N} \  \biggl(\frac{M}{E}\biggr)^{2N}
        \ \ \ \ \ \ 
        N\ \mbox{even}.
\label{eff:est_even}
\end{eqnarray}
Here the first factor comes from the $N$ loop integrals and $N$ gauge boson
propagators while the second factor comes from two interfering currents to
radiate $N$ gauge bosons.

The case of odd $N$ is treated analogously. The Feynman rule corresponding
to radiation of $N$ gauge bosons from decay is
\be
N\ \ \ 
\begin{picture}(60,20)(0,-2) 
        \GCirc(30,0){3}{0}
        \ArrowLine(30,0)(60,10)
        \ArrowLine(60,-10)(30,0)
        \Photon(30,0)(2,10){2}{6}
        \Photon(30,0)(2,-10){2}{6}
        \Photon(30,0)(0,4){2}{6}
        \Photon(30,0)(0,-4){2}{6}
        \DashCArc(30,0)(35,-20,20){1}
        \DashCArc(30,0)(35,160,200){1}
\end{picture}
        \ \ \ 
        (P_M,p)
        \ \sim \ \ 
        \DD \ 
        \frac{e^N}{\Lt^N}\frac{(P_M^2)^{(N+1)/2}}{\Lo^{N+1}},
        \ \ \ \ \ \ 
        N\ \mbox{odd}.
\ee
In this case, the relative $N$-loop non-factorizable correction is
\begin{eqnarray}
        \delta_N
&\sim&
        \int \frac{d^4k_1}{k_1^2}\ldots \frac{d^4k_N}{k_N^2} \ 
        \biggl(\frac{e^N}{\Lt^N}\frac{(P_M^2)^{(N+1)/2}}{\Lo^{N+1}}\biggr) \ 
        \biggl(\frac{e^N}{\Lt^N}\frac{(P_M^2)^{(N+1)/2}}{\Lo^{N+1}}\biggr)
\nonumber \\
&\sim&
        \biggl(\frac{\Lt^4}{\Lt^2}\biggr)^N \ 
        \biggl(\frac{e^N}{\Lt^N}\frac{(P_M^2)^{(N+1)/2}}{\Lo^{N+1}}\biggr)^2 \ 
        \sim e^{2N} \  \biggl(\frac{M}{E}\biggr)^{2(N+1)}
        \ \ \ \ \ \ 
        N\ \mbox{odd}.
\label{eff:est_odd}
\end{eqnarray}
Thus, the estimates obtained using
the effective theory, \Eqns{eff:est_even}{eff:est_odd}, agree with the
estimates obtained through the analysis of higher order Feynman
diagrams,  \Eqns{feyn:est_even}{feyn:est_odd}.


\section{Applications}
\label{sec:appl}

In this section we consider some topical applications of the
techniques presented in Sect.~\ref{sec:feyn} and Sect.~\ref{sec:eff}
to the pair production of $W$ and $Z$ bosons and top quarks
in $\gamma\gamma$ and $e^+e^-$ collisions.%
\footnote{Our approach may be generalized to the case of
  pair production of unstable particles in hadronic reactions (for
  discussion of the gluon bremsstrahlung effects accompanying the
  $t\bar{t}$ production at hadron colliders see for instance
  \cite{kos1}).}
As above, all our considerations will be performed in the pole-scheme.
We shall also confine ourselves to the case when no kinematic
restrictions are imposed on the configurations of the final state
particles.  In general, as shown in \cite{cks-rad}, the kinematic cuts
applied to the final state particles may induce significant changes in
the energy behavior of the non-factorizable effects.  We focus on the
distributions in which the integration is performed over the momenta
of the decay products keeping tracks of the invariant masses of the
unstable particles.

In Sect.~\ref{sec:feyn} and Sect.~\ref{sec:eff} it was shown that
non-factorizable corrections due to interaction with decay dipoles are
suppressed at high energy.  Throughout this section we consider only
this type of corrections.  We do not discuss other corrections
(factorizable corrections and non-factorizable corrections due to
propagation corrections) for which the suppression mechanism is not
applicable.

\subsection{Neutral unstable particles}

We begin with the basic case when only neutral (both with respect to
electromagnetic and color charges) particles participate in the
production process and, therefore, only interference between
radiation accompanying the decays occurs. This can be exemplified by
the process
\be
\label{appl/zz/aa->4q}
	\gamma\gamma \to Z(p_1) Z(p_2)\to 4 \ \mbox{fermions} \
	(k_1,k_1';k_2,k_2'), 
\ee
with $p_{1,2}=k_{1,2}+k_{1,2}'$.  The non-factorizable effects appear
in the first order in QED and in the second order in QCD.  In the case
of color singlet unstable particles the QCD interconnection is
additionally suppressed by $1/N_c^2$ compared to the leading QCD
contribution, see e.g. \cite{sjos-nonpert}
($N_c$ being the number of colors).

For the purposes of illustration it is sufficient to focus attention
on the pure QED phenomena accompanying the process
\be
\label{appl/zz/aa->4l}
	\gamma\gamma \to Z(p_1) Z(p_2)\to
	e^+(k_1)e^-(k_1')\mu^+(k_2)\mu^-(k_2'). 
\ee
We present below an estimate of the energy behavior of the
non-factorizable correction of order $\OO(\alpha^N)$ induced by the
exchanges of $N$ virtual photons with momenta $l_1,..,l_N$ between the
decay products of the two $Z$'s.

Similarly to what we did in Sect.~\ref{sec:feyn} in order to find the
$N$-loop contribution we have to evaluate the integrals over the Born
decay matrix element squared multiplied by the $N$ non-factorizable
currents corresponding to radiation off the massless charged fermions
with momenta $k_{1,2}$ and $k_{1,2}'$ keeping the momenta of the $Z$
bosons, $p_{1,2}$, fixed
\bea
	I^{\alpha_1\ldots\alpha_N}_{\mu\nu}
	&=&
	\int d^4k_i \ d^4k_i'  \ 
	\delta(k_i^2) \ \delta(k_i^{\prime 2}) \ 
	\delta^4(p_i-k_i-k_i') \ \times \ 
	\Delta_{\mu\nu}
\nonumber \\
\label{appl/zz/int}
  &\times&
	\biggl(\frac{k_i}{k_il_1}-\frac{k_i'}{k_i'l_1}\biggr)^{\alpha_1}
	\ldots
	\biggl(\frac{k_i}{k_il_N}-\frac{k_i'}{k_i'l_N}\biggr)^{\alpha_N},
\eea
where $i=1,2$.

The new element with respect to the consideration of
Sect.~\ref{sec:feyn} is the appearance of the decay tensor
$\Delta_{\mu\nu}$, which results from the vector nature of the
decaying particles. The indices $\mu$, $\nu$ will be contracted with the
rest of the Born matrix element squared. The indices
$\alpha_1\ldots\alpha_N$ will be contracted with the corresponding
indices of the non-factorizable currents from the second unstable
particle.  $\Delta_{\mu\nu}$ can be decomposed into  two
contributions with  different $P$-parity
\be
	\Delta_{\mu\nu} = \Delta_{\mu\nu}^{V} + \Delta_{\mu\nu}^{A} 
\ee
\be
	\Delta_{\mu\nu}^{V} \sim k_\mu k_\nu' + k_\mu' k_\nu - (kk') g_{\mu\nu},
	\ \ \ 
	\Delta_{\mu\nu}^{A} \sim \epsilon_{\mu\nu \sigma \rho} k^\sigma k^{\prime \rho}.
\ee
The $\Delta_{\mu\nu}^{V}$ piece corresponds to the \peven\
contribution to the matrix element squared while
$\Delta_{\mu\nu}^{A}$ results from the \podd\ interference.  We shall
call them \peven\ and \podd\ contributions respectively.

Similarly to Sect.~\ref{sec:feyn} we introduce the new variables
\be
\label{appl/zz/Delta}
	k_i^\mu = \frac{1}{2}\bigl(p_i^\mu+\Delta_i^\mu\bigr),
	\ \ \ 
	k_i^{\prime \mu} = \frac{1}{2}\bigl(p_i^\mu-\Delta_i^\mu\bigr).
\ee
In terms of these new variables the decay tensor components
$\Delta_{\mu\nu}^{V,A}$ can be rewritten as
\be
\label{appl/zz/born}
	\Delta_{\mu\nu}^{V} 
	= 
	\frac{1}{2} \biggl[p^\mu p^\nu-M^2 g^{\mu\nu}-\Delta^\mu\Delta^\nu\biggr],
	\ \ \ 
	\Delta_{\mu\nu}^{A}
	=
	\frac{1}{2}
	\epsilon_{\mu\nu \sigma \rho} \Delta^\sigma p^{\rho}.
\ee
We immediately see from Eq.~(\ref{appl/zz/born}) that the
$\Delta$-dependent part of the \peven\ component is suppressed by a
factor $\sim M^2/E^2$ as compared to the leading term while the
\podd\ component is suppressed by $\OO(M/E)$ factor with respect to the
leading \peven\ part.

Let us perform an integration over the final state particle momenta in
\Eqn{appl/zz/int} (see Sect.~\ref{sec:feyn} and, in particular,
Eq.~(\ref{dec/int}) and Eq.~(\ref{dec/delta_int})) and evaluate the
high-energy behavior of different components of the $N$-loop tensor
$I^{\alpha_1\ldots\alpha_N}_{\mu\nu}$
\be
	{}^{V(A)} I_{\mu\nu}^{\alpha_1\ldots\alpha_N}
	=
	\int_\Delta \Delta_{\mu\nu}^{V(A)}
	\cdot
	\biggl(\frac{k}{kl_1}-\frac{k'}{k'l_1}\biggr)^{\alpha_1}
	\ldots
	\biggl(\frac{k}{kl_N}-\frac{k'}{k'l_N}\biggr)^{\alpha_N}.
\ee
The result depends on whether $N$ is odd or even
\be
\label{appl/zz/est-v}
	{}^V I_{\mu\nu}^{\alpha_1\ldots\alpha_N}
	\sim
	\left\{
	\begin{array}{l}
	0,  \ \ \ \mbox{for $N$ \ odd},\\
	\mbox{Born} \cdot \Gamma^{-N}, \ \ \  \mbox{for $N$ \ even}.
	\end{array}
	\right. 
\ee
\be
\label{appl/zz/est-a}
	{}^A I_{\mu\nu}^{\alpha_1\ldots\alpha_N}
	\sim
	\left\{
	\begin{array}{l}
	\mbox{Born} \cdot \Gamma^{-N}\cdot M/E,  \ \ \ \mbox{for $N$ \ odd},\\
	0, \ \ \  \mbox{for $N$ \ even}.
	\end{array}
	\right. 
\ee 
Here ``Born'' is a symbolic notation for the leading \peven\
contribution to the Born matrix element squared.

Power-counting rules established in Sect.~\ref{sec:feyn} allow one to
estimate the suppression factors corresponding to the different parts
of the non-factorizable correction to the cross section of the process
(\ref{appl/zz/aa->4l}). The results are summarized in
Table~\ref{tab:appl/zz} where the notations \peven\ -- \peven, \peven\
-- \podd, \podd\ -- \podd\ are used to specify the interference between
the different contributions into the Born decay matrix element squared
of each $Z$-boson.
\begin{table}[ht]
\[
  \begin{array}{||c||c|c|c||}    \hline \hline
    \raisebox{-3mm}{N} & \multicolumn{3}{|c||}{\raisebox{-1mm}{Born}}\\ 
        \cline{2-4}
                       & \sss{\peven\ -- \peven} &  \sss{\peven\ -- \podd}  & \sss{\podd\ -- \podd}  \\
    \hline  \hline
    N\mbox{-even}      & (M/E)^{2N}      & 0 & 0  \\   
    \hline   
    N\mbox{-odd}       & 0               & 0 & (M/E)^{2N+2} \\
    \hline \hline
  \end{array}
\]
\ccaption{}{Estimates of the energy dependence of the 
		$\OO(\alpha^N)$ non-factorizable correction,
		$\delta_{\sss{nf}}$, integrated over the momenta of
		the decay products for different interference
		contributions corresponding to the process
		(\ref{appl/zz/aa->4l}).
\label{tab:appl/zz} }
\end{table}%

As follows from Table~\ref{tab:appl/zz}, the one-loop QED
non-factorizable corrections to the cross section of the process
(\ref{appl/zz/aa->4l}) for different final states may appear only as a
result of the \podd\ -- \podd\ decay interference and in practice are
numerically small (see Ref.~\cite{nf-zz}).  This term scales as
$E^{-4}$ with the CMS energy, which is in accord with the earlier
observation in Ref.~\cite{nf-anz}. The two-loop QED interconnection
effects are caused by the interference of the $P$-even terms \peven\
-- \peven, and their contribution also scales as $E^{-4}$.
Symbolically, the four leading QED terms in relative non-factorizable
correction, $\delta_{\sss{nf}}$, for the process
(\ref{appl/zz/aa->4l}) can be presented as
\be
\label{appl/zz/est-qed}
	\delta_{\sss{nf}} \ \sim \   
	r_{A}\ 
	\alpha  \biggl(\frac{M}{E}\biggr)^4
	\biggl[1+\alpha^2\biggl(\frac{M}{E}\biggr)^4+\ldots\biggr]
	+
	\alpha^2 \biggl(\frac{M}{E}\biggr)^4
	\biggl[1+\alpha^2\biggl(\frac{M}{E}\biggr)^4+\ldots\biggr],
\ee
where parameter $r_A$ is determined by the product of vector and axial
couplings of the corresponding neutral current.  Therefore, the actual
parameter of the perturbation theory in $\delta_{\sss{nf}}$ separately
for even and odd powers of $\alpha$ is $\alpha^2 (M/E)^4$, and the
series in \Eqn{appl/zz/est-qed} rapidly converges at $E\gg M$.

All these observations remain valid for the decay-decay part of the
QED non-factorizable corrections to the process
\be
\label{appl/zz/ee->4q}
	e^+e^- \to Z(p_1) Z(p_2)\to 4 \ \mbox{fermions} \ (k_1,k_1';k_2,k_2'),
\ee
The difference is that now there are also production-decay QED
interferences, which were absent in the process
(\ref{appl/zz/aa->4l}). Estimates for production-decay interferences
are in general different from those for the decay-decay interferences.

Let us now turn to the QCD non-factorizable corrections to the
processes (\ref{appl/zz/aa->4q}) and (\ref{appl/zz/ee->4q}) with final
state quarks.  Due to the group structure of QCD these can appear
only at $N\geq 2$, see e.g. \cite{sjos-nonpert}.  Only the \peven\ --
\peven\ decay-decay gluon interference survives at $N=2$ with the
energy scaling $E^{-4}$. As we discussed in
Sect.~\ref{sec:eff/resonant_neutral/matching/QCD}, in QCD the
separation between higher-order non-factorizable corrections due to
interaction with decay dipoles, which is of interest to us here, and
non-factorizable propagation corrections is complicated due to the
non-abelian structure of the gauge group.  This separation is
correctly understood in terms of different operators in the Lagrangian
of the effective theory discussed in
Sect.~\ref{sec:eff/resonant_neutral/matching/QCD}.  What we estimate
here are the leading non-factorizable corrections due to operators
describing decay dipole interactions.  Using Table~\ref{tab:appl/zz}
we can see that the four leading terms in $\delta_{\sss{nf}}^{QCD}$
can be written symbolically as
\be
	\label{appl/zz/est-qcd}
	\delta_{\sss{nf}}^{QCD} \ \sim \   
	\alpha_{\sss{S}}^2 \biggl(\frac{M}{E}\biggr)^4
	\biggl[1+ \alpha_{\sss{S}}^2\biggl(\frac{M}{E}\biggr)^4+\ldots\biggr]
	+
	r_A 
	\alpha_{\sss{S}}^3  \biggl(\frac{M}{E}\biggr)^8
	\biggl[1+ \alpha_{\sss{S}}^2\biggl(\frac{M}{E}\biggr)^4+\ldots\biggr].
\ee
The $N_c$ dependence of non-factorizable corrections might be of
special interest. The exact formula is complicated for arbitrary
order, $N$, however, the leading $N_c$ dependence is easy to estimate
and goes as $(C_F \alpha_{\rm S})^N/N_c^2 \sim (N_c\alpha_{\rm
S})^N/N_c^2$.  This means that the actual parameter of the
perturbation theory separately for even and odd powers of
$\alpha_{\sss{S}}$ is $N_c^2 \alpha_{\sss{S}}^2 (M/E)^4$.  The same
conclusion remains valid for the QCD non-factorizable corrections to
the process
\be
	e^+e^-[\gamma\gamma] \to W^+ W^- \to 4 \ \mbox{quarks}.
\ee
The main qualitative difference between the $W^+W^-$ and $ZZ$
interconnection results stems from the difference in the vector and
axial couplings of the $W$ and $Z$ bosons.  Note that since the $Z$
mass and other properties are well known, $ZZ$ production provides a
unique laboratory for studying QCD interconnection
phenomena~\cite{sjos-nonpert, sjos-nonpert-x}.

Formally, the application of perturbation theory to the description of
non-factorizable  QCD phenomena accompanying $ZZ$ and $W^+W^-$
production at $E\gg M$ is restricted by the requirement that the
typical space-time separation between the decay vertices does not
exceed the characteristic strong interaction scale $R_{\sss{ch}}\sim
1/\mu$, where $\mu\sim m_{\pi}\sim \Lambda_{QCD}$,
\be
\label{appl/zz/pert_condition}
	d\sim 1/\Gamma E/M < 1/\mu,
\ee
see Refs.~\cite{kos-rad,sjos-nonpert}.  In other words, remembering
that the life-time of an unstable particle in the laboratory frame is
\be
	t_{\sss{dec}} \sim \frac{E}{M} \frac{1}{\Gamma},
\ee
we can say that this condition implies that the decay happens before
the formation of the first interconnecting hadrons.

At the same time, since the relevant momentum scale for the
non-factorizable effects is $k\sim 1/d$ the requirement
\Eqn{appl/zz/pert_condition} coincides with the Landau pole
condition for $\alpha_{\sss{S}}$ associated with the
$\delta_{\sss{nf}}^{QCD}$ correction.  Therefore, even though formally
the perturbative results at very high energies are not justified our
estimates show that in the non-controllable domain the
$\delta_{\sss{nf}}^{QCD}$ corrections due to interaction with products
of decay are decreasing very sharply with increasing energy, if the two
outgoing hadronic systems are more and more boosted apart.  Moreover,
the higher is the power of $\alpha_{\sss{S}}$ coupling to the decay
dipoles the steeper is the energy fall-off.  The corrections to
propagation, on the other hand, are not suppressed at high energy.  At
large $\alpha_{\sss{S}}$ the description of propagation becomes purely
non-perturbative.  In is worthwhile to mention that the
non-perturbative models predict that the hadronic cross-talk between
the decaying particles dampens with energy comparatively slowly
\cite{sjos-nonpert-x,khoze-sjos}.

\subsection{Charged unstable particles}

In the case of charged, unstable particles both types of
non-factorizable corrections are present: decay-decay and
production-decay.  We shall consider only the decay-decay part.  Our
analysis embraces, in particular, QED effects in the $W^+W^-$
production process and QCD effects in the production and decay
of a $t\bar{t}$ pair
\be
	t\bar{t} \to b W^+ \bar{b} W^-.
\ee
Let us consider the case when the ``charged'' (with respect to QED or
QCD interactions as appropriate) unstable particle with momentum
$p^\mu$ decays into a ``charged'' massless particle with momentum
$k^\mu$ and a ``neutral'' particle with momentum $k^{\prime\mu}$.  We
begin with vector-boson pair production accompanied by the
leptonic decays.  Analogously to the previous subsection we introduce
the new variable $\Delta$ (see Eq.~(\ref{appl/zz/Delta})) and evaluate
the high-energy behavior of the $N$-loop tensors ${}^{V(A)}
I_{\mu\nu}^{\alpha_1\ldots\alpha_N}$
\be
\label{appl/ww/int}
	{}^{V(A)} I_{\mu\nu}^{\alpha_1\ldots\alpha_N}
	=
	\int_\Delta \Delta_{\mu\nu}^{V(A)}
	\cdot
	\biggl(\frac{p}{pl_1}-\frac{k}{kl_1}\biggr)^{\alpha_1}
	\ldots
	\biggl(\frac{p}{pl_N}-\frac{k}{kl_N}\biggr)^{\alpha_N}.
\ee
Here the contributions with different $P$-parity are determined by the
Born decay matrix elements.  Repeating the same procedure as in the
derivation of Eqs.~(\ref{appl/zz/est-v}) and (\ref{appl/zz/est-a}) we
arrive at
\be
\label{appl/ww/est-v}
	{}^V I_{\mu\nu}^{\alpha_1\ldots\alpha_N}
	\sim
	\left\{
	\begin{array}{l}
	\mbox{Born} \cdot \Gamma^{-N} \cdot M/E,  \ \ \ \mbox{for odd $N$},\\
	\mbox{Born} \cdot \Gamma^{-N}, \ \ \  \mbox{for even $N$}.
	\end{array}
	\right. 
\ee
\be
\label{appl/ww/est-a}
	{}^A I_{\mu\nu}^{\alpha_1\ldots\alpha_N}
	\sim
	\left\{
	\begin{array}{l}
	\mbox{Born} \cdot \Gamma^{-N}\cdot M/E,  \ \ \ \mbox{for odd $N$},\\
	\mbox{Born} \cdot \Gamma^{-N}\cdot M^2/E^2, \ \ \  \mbox{for even $N$}.
	\end{array}
	\right. 
\ee 
Again, ``Born'' is a symbolic notation for the leading \peven\
contribution to the Born matrix element squared.  Recall that the main
difference with respect to the case of neutral particles is caused by
the presence of the odd powers of $\Delta$ in the expansion of the
decay currents over $\Delta$.

The estimates for the energy scaling of the non-factorizable
corrections caused by the final-final interferences are summarized in
Table~\ref{tab:appl/ww}.
\begin{table}[ht]
\[
  \begin{array}{||c||c|c|c||}    \hline \hline
    \raisebox{-3mm}{$N_1, N_2$} &
    \multicolumn{3}{|c||}{\raisebox{-1mm}{Born}}\\  
        \cline{2-4}
                       	& \sss{\peven\ -- \peven} & \sss{\peven\ -- \podd} & \sss{\podd\ -- \podd}\\
    \hline  \hline
    N \mbox{-even}	& (M/E)^{2N} 	& (M/E)^{2N+2} & (M/E)^{2N+4}  \\   
     \hline   
    N\mbox{-odd}	& (M/E)^{2N+2} 	& (M/E)^{2N+2} & (M/E)^{2N+2}  \\ 
    \hline \hline
  \end{array}
\]
\ccaption{}{Estimates of the energy dependence of the $\OO(\alpha^N)$
    non-factorizable corrections, integrated over the momenta of the
    decay products for the decay-decay interference contributions
    associated with the ``charged'' unstable particle production.
\label{tab:appl/ww}}
\end{table}%

From Table~\ref{tab:appl/ww} we can see immediately  that the leading
part of the decay-decay interference contribution to
$\delta_{\sss{nf}}^{QED}$ in the case of the $W^+W^-$ production
scales with energy as $E^{-4}$. This was first established in
Ref.~\cite{nf-anz}.  Symbolically the four leading QED terms for this
process can be written as
\be
\label{appl/ww/est-qed}
	\delta_{\sss{nf}} \ \sim \   
	\alpha  \biggl(\frac{M}{E}\biggr)^4
	\biggl[1+\alpha^2\biggl(\frac{M}{E}\biggr)^4+\ldots\biggr]
	+
	\alpha^2 \biggl(\frac{M}{E}\biggr)^4
	\biggl[1+\alpha^2\biggl(\frac{M}{E}\biggr)^4+\ldots\biggr],
\ee

A similar argumentation remains valid for the QCD non-factorizable
corrections corresponding to the $t\bar{t}$ production process
\be
\label{appl/ww/aa->tt}
	\gamma\gamma(e^+e^-) \to t\bar{t}\to  \ b W^+ \bar{b} W^-,
\ee	
with $\alpha$ substituted by $\alpha_{\sss{S}}$.  Recall that in the
$t\bar{t}$ case there is no color suppression in
$\delta_{\sss{nf}}^{QCD}$ if the pair is produced in the color
singlet state and there is an additional $1/N_c^2$ suppression for the
case of the octet $t\bar{t}$ state~\cite{kos1,nf-tt}.
Analogously to the case of color-singlet unstable particle production
discussed above the perturbative results for $\delta_{\sss{nf}}^{QCD}$
are fully controllable only if the restriction
\Eqn{appl/zz/pert_condition} is satisfied.

We stress again that for the charged particles case this estimates are
not complete.  For the evaluation of the magnitude of the total QCD
interconnection effects one has to incorporate also production-decay
interferences and real radiation.


\section{Conclusions}
\label{sec:concl}

In the near future, the direct evaluation of two-loop corrections to
processes involving unstable particles is not feasible, not to mention
even higher order corrections. In view of the phenomenological
importance of these processes, however, it is essential to improve our
understanding beyond the presently available one-loop calculations. In
this paper we took a first step in this direction by analyzing
higher-order radiative corrections to pair production of unstable
particles with their subsequent decay.

Having a future linear collider in mind, we are particularly
interested in the high-energy behavior, $E\gg M$, of the
corrections. We identified various effects and studied their
behavior/suppression at high energy. In particular, we considered
distributions that are obtained by integrating over the angles of the
decay products, but keeping their invariant mass fixed.

As a first case, we studied neutral, unstable particles decaying into
fermions that interact through the exchange of abelian gauge
bosons. This is the simplest case and can be analyzed by studying
Feynman diagrams. By extending the well known one-loop
analyses~\cite{dpa-calc, nf-calc/bbc, nf-calc/ddr}, we found that to
all orders in perturbation theory the complete corrections have the
following structure: there are factorizable and non-factorizable
corrections and interferences between them that factorize with respect
to each other, \Eqn{feyn/Nloop/full_struct}. The factorizable
corrections are associated with a hard scale $\sim E$, whereas the
non-factorizable corrections are associated with a soft scale $\sim
\Gamma M/E$. Furthermore, there are two sources of non-factorizable
corrections, namely interactions of the gauge bosons with the decay
dipoles and propagator corrections of the gauge bosons.

Since the various corrections listed above have a different physical
origin it is to be expected that their high-energy behavior is
different. Indeed, we found that the corrections due to the
interaction with the dipoles is --- on top of the usual coupling
constant suppression --- additionally suppressed at high energies,
\Eqns{feyn:nfneutraleven}{feyn:nfneutralodd}. The other
corrections identified above do not have such an additional
suppression. This analysis can be generalized to the case of charged,
unstable particles and similar results are found,
\Eqns{feyn:est_even}{feyn:est_odd}. 

The extension of such an analysis to the non-abelian case is a rather
daunting task. The reason is that the self-coupling of the gauge
bosons complicates the matter enormously. In particular, most diagrams
will now contribute to several types of corrections, rather than just
one. Therefore, it was necessary to tackle this problem in a
different way. Since the separation of the full corrections
into various parts relies on the presence of two different scales,
$\Lo\sim E$, $\Lt\sim\Gamma M/E$ with $\Lo\gg\Lt$, an effective
theory approach looks very promising.

In order to pursue this we identified the relevant fields present in
the underlying theory: hard, resonant, soft and external. By
subsequently integrating out the various fields we obtain a hierarchy
of effective field theories. The first step is to integrate out the
hard modes. This leaves us with a theory where only resonant, soft and
external particles are dynamical. The factorizable corrections
are those that are encoded in the (higher order corrections to the)
Wilson coefficients, whereas the non-factorizable corrections are the
corrections due to the still dynamical degrees of freedom. What we
have gained is that the generalization of this separation to higher
orders in the coupling constant and to the non-abelian case is
straightforward. Throughout the paper we restrict ourselves to the
leading terms in $\Lt/\Lo$. In this approximation, the effective
theory is equivalent to the DPA of the underlying theory.

The second step in constructing an effective field theory is to
integrate out the resonant modes. The corresponding effective
Lagrangian contains again (non-local) gauge-invariant operators,
multiplied by Wilson coefficients. Some of these operators are
responsible for the propagator corrections and some are responsible
for the dipole interactions. Thus, in the same way as integrating out
hard modes provided us with a separation of factorizable and
non-factorizable interactions, integrating out resonant modes allows
us to separate the two kinds of non-factorizable corrections. As a
consequence of the gauge invariance of the operators, this separation
is also gauge invariant.

In order to obtain estimates of the various corrections similar to
those obtained by analyzing Feynman diagrams, all we have to do is to
get estimates of the Wilson coefficients of the corresponding
operators. These estimates can be obtained by general considerations,
i.e. without explicitly knowing the Wilson coefficient. Thereby we
generalize the estimates obtained in the abelian case to the
non-abelian case. Obviously, if we wanted to do a precise calculation
in the effective theory we would need the exact form of the Wilson
coefficient. They have to be obtained through matching the effective
theory to the underlying theory. We have done this matching for the
operators corresponding to the emission of up to three gauge bosons,
thereby confirming the estimates obtained by general considerations.

We applied the estimates to several phenomenologically relevant
processes. The actual estimate depends on the structure of the Born
term. Also, it is important to stress again that the estimates
obtained here are valid only for the non-factorizable corrections due
to dipole interactions for distributions integrated over the decay
angles. Furthermore, in the effective theory approach we considered
the case of neutral, unstable particles only.

In the case of charged, unstable particles there are additional
non-factorizable corrections of the production-decay type. In the
effective theory language this means that the operator associated with
$\PP$, \Eqn{eff/fullP}, is more complicated than given in
\Eqn{eff/simpleP}. Allowing for the emission of gauge bosons from the
unstable particle production process results in more complicated
operators in the effective Lagrangian. In order to obtain estimates in
this case one would have to construct these operators and obtain
estimates of the corresponding Wilson coefficients. 

Another possible future development is the inclusion of real
gauge-boson radiation. As mentioned in the introduction, real
gauge-boson radiation contributes to the non-factorizable
corrections. In this paper, we have left this issue completely
aside. However, adding real radiation and performing an integration
over the phase space could certainly be studied in the effective
theory framework.

Finally, we would like to mention the possibility to extend this
analysis beyond leading order in $\Lt/\Lo$. As mentioned above, if we
keep only the leading terms in $\Lt/\Lo$ after integrating out the
hard modes we recover the DPA of the underlying theory. Since there is
nothing that prevents us to go beyond the leading terms in $\Lt/\Lo$,
the effective theory approach opens up the possibility to go beyond
the DPA in a systematic way.

In summary, we studied the structure of higher-order corrections in
the presence of unstable particles, exploring an effective field
theory approach to the problem. Even though we made several
simplifying restrictions this approach is very promising and seems to
allow to look at a whole set of problems from a different, more
promising, point of view.

\subsection*{Acknowledgements}
VAK thanks the Leverhulme Trust for a Fellowship. Work supported in
part by the EU Fifth Framework Programme `Improving Human Potential',
Research Training Network `Particle Physics Phenomenology at High
Energy Colliders', contract HPRN-CT-2000-00149.


\setcounter{equation}{0}
\def\theequation{A\arabic{equation}}
\setcounter{section}{0}
\def\thesection{\Alph{section}}

\section{Expansion of loop integrals \label{app:a}}

In this appendix we consider an example of the separation of
factorizable and non-factorizable corrections to all orders in
$\Lambda_2/\Lambda_1$. We use an expansion of loop integrals in
dimensional regularization by letting the loop momenta be in all
relevant momentum regions \cite{beneke-smirnov, smir00}. This
expansion is equivalent to our effective field theory framework.

The diagram we consider here is shown in
Fig.~\ref{fig:eff/hard/corr}(c) and contains both factorizable and
non-factorizable contributions. The scalar integral associated with
this diagram is
\be
	I=
	\int 
	\frac{d l}{l^2}\frac{1}{(l^2-2kl)(l^2-2pl+D)},
\ee
where $k^2=m^2$, $p^2=M^2$, $M\gg D\gg m$.  The integral is IR and UV
finite and, thus, does not require regularization.  The exact result
is
\be
	I = \frac{i\pi^2}{M^2}
	\biggl[
	\ln\frac{M^2}{m^2}\ln\frac{-D}{M^2-D} 
	-\Li\biggl(\frac{D}{M^2}\biggr)
	-\frac{1}{2}\ln^2\frac{M^2-D}{M^2}
	-\frac{\pi^2}{6}
	\biggr].
\ee

In order to expand the loop integral in $D/M^2$ we should assume the
loop momentum to be in one of the following two regions: soft, $l\sim
D/M$, and hard, $l\sim M$. Then we have to expand the integrand and
perform the integrations over the full momentum space in dimensional
regularization.
\be
	I = I_{\sss{soft}} + I_{\sss{hard}}
	=
	\int 
	\frac{d l}{l^2}\frac{1}{(-2kl)(-2pl+D)}
	+
	\int 
	\frac{d l}{l^2}\frac{1}{(l^2-2kl)(l^2-2pl)}
	\sum\limits_{n=0}^{\infty}\biggl(\frac{-D}{l^2-2pl}\biggr)^n.
\ee
There is only one (the leading) term in the soft momentum expansion
because in higher orders the $l^2$ term in the denominator will be
cancelled. The resulting integrals are scaleless and vanish in
dimensional regularization. The soft integral is UV divergent, and the
hard integral is IR divergent. These divergences should be regularized
by dimensional regularization and they will cancel in the sum.

The soft integral is 
\be
	I_{\sss{soft}}
	=
	-\frac{i\pi^2}{M^2}
	\biggl[
 \ln\frac{M}{m}\biggl(\frac{1}{\epsilon}-\gamma-
 \ln\pi+2\ln\frac{M^2}{-D}-\ln M^2\biggr)
	+\ln^2\frac{M}{m} + \frac{\pi^2}{6}
	\biggr].
\ee
For  $n=0$ the hard integral is given by
\be
	\int 
	\frac{d l}{l^2}\frac{1}{(l^2-2kl)(l^2-2pl)}
	=
	\frac{i\pi^2}{M^2}
	\biggl[
	\ln\frac{M}{m}\biggl(\frac{1}{\epsilon}-\gamma-\ln\pi-\ln M^2\biggr)
	+\ln^2\frac{M}{m}
	\biggr].
\ee
whereas for  $n\geq1$ the integrals are finite and given by
\bea
	\lefteqn{\int 
	\frac{d l}{l^2}\frac{1}{(l^2-2kl)(l^2-2pl)}
	\sum\limits_{n=1}^{\infty}\biggl(\frac{-D}{l^2-2pl}\biggr)^n
	=} &&
\\ \nonumber &&
	\frac{i\pi^2}{M^2}
	\biggl[
	-\ln\biggl(1-\frac{D}{M^2}\biggr)\ln\frac{M^2}{m^2}
	-\Li\biggl(\frac{D}{M^2}\biggr)
	-\frac{1}{2}\ln^2\biggl(1-\frac{D}{M^2}\biggr)
	\biggr].
\eea
As expected, the sum of $I_{\sss{soft}}$ and  $I_{\sss{hard}}$
reproduce the full result.

\section{Matching calculation \label{app:b}}

\setcounter{equation}{0}
\def\theequation{B\arabic{equation}}

In this appendix we give some details concerning the matching of the
$(\bar{\psi}\psi) F^3$ operator defined in \Eqn{eff/operator_3F/QCD}.
To do this matching, we first have to evaluate the contribution of the
operator $(\bar{\psi}\psi) F^2$, \Eqn{eff/operator_2F/QCD}, to
three-gluon processes. The Wilson coefficient
$\Sigma^{(3)}_{\mu_1\nu_1\mu_2\nu_2\mu_3\nu_3}(k,k'|l_1,l_2,l_3)$ is
then determined such that the full result in the effective theory is
the same as in QCD.

The operator $(\bar{\psi}\psi) F^2$ yields two different contributions
to three-gluon processes. The first is obtained by taking the leading
term in the path ordered exponentials, $U(z_1,z_2) = 1$, while taking
the abelian part of one of the two field tensors $F_{\mu\nu}$ and the
commutator term of the other. The evaluation of these terms is
straightforward. The second contribution is obtained by taking the
abelian part in both field tensors but the order $g$ term in one path
ordered exponential. In this case we have to evaluate expressions of
the form
\bea
&&
\big(l_1^{\mu_1}g^{\nu_1\alpha_1}-l_1^{\nu_1}g^{\mu_1\alpha_1} \big)
\big(l_2^{\mu_2}g^{\nu_2\alpha_2}-l_2^{\nu_2}g^{\mu_2\alpha_2} \big)
(\Tr(T^{a_1}T^{a_2}T^{a_3})-\Tr(T^{a_1}T^{a_3}T^{a_2}))
\phantom{\int}
\label{sig2} \\
&&
\int dx dy d^2z d^4s \
e^{is_1(x-X)}e^{is_2(y-X)}e^{is_3(z_1-X)}e^{is_4(z_2-X)}
e^{-ikx}e^{-ik'y}e^{-il_1z_1}e^{-il_2z_2}
\nonumber \\
&&
\Sigma^{(2)}_{\mu_1\nu_1\mu_2\nu_2}(s_1,s_2|s_3,s_4) (-ig)
\int_{z_1}^{z_2}d\omega^{\alpha_3}\ e^{-il_3\omega} 
\quad +\ \{1\leftrightarrow 2\leftrightarrow 3 \}
\nonumber
\eea
where $l_i, \alpha_i$ and $a_i$ are the momenta, Lorentz indices and
color labels of the gluons respectively.  The first point to note is
that this contribution depends on the chosen path from $z_1$ to $z_2$.
This path dependence has to be cancelled by the contribution of the
operator $(\bar{\psi}\psi) F^3$. From this we conclude that the
precise form of the corresponding Wilson coefficient $\Sigma^{(3)}$ is
path dependent.

To illustrate this point we will consider the matching for two
different paths. It is convenient to split the tensor $\Sigma^{(3)}$
into a color symmetric and antisymmetric piece
\be
  \Sigma^{(3)}_{\mu_1\nu_1\mu_2\nu_2\mu_3\nu_3}(k,k'|l_1,l_2,l_3) = 
  \Sigma^{(3;+)}_{\mu_1\nu_1\mu_2\nu_2\mu_3\nu_3}(k,k'|l_1,l_2,l_3) + 
  \Sigma^{(3;-)}_{\mu_1\nu_1\mu_2\nu_2\mu_3\nu_3}(k,k'|l_1,l_2,l_3)
\ee
The tensor $\Sigma^{(3;+)}$ will be proportional to a color factor
$1/2\ (\Tr(T^{a_1}T^{a_2}T^{a_3})+\Tr(T^{a_1}T^{a_3}T^{a_2})) $ while
the $\Sigma^{(3;-)}$ terms will come with a color factor $1/2\
(\Tr(T^{a_1}T^{a_2}T^{a_3})-\Tr(T^{a_1}T^{a_3}T^{a_2}))$. The
symmetric part is exactly the same as in the abelian theory and is
given in \Eqn{wilson_abelian} with $N=3$. The antisymmetric part is
the new feature of the non-abelian case and contains all the path
dependence.

The first path we consider is the linear path. In this case
\be
\int_{z_1}^{z_2}d\omega^{\alpha_3}\ e^{-il_3\omega} =
i\, \frac{z_2^{\alpha_3}-z_1^{\alpha_3}}{l_3\cdot (z_2-z_1)} 
\left( e^{-il_3z_2} - e^{-il_3z_1} \right)
\ee
Thus the contribution of the operator $(\bar{\psi}\psi) F^2$ to
three-gluon processes given in \Eqn{sig2} will contain logarithms with
ratios of scalar products of momenta as arguments. Since the QCD
result does not contain such logarithms, they will have to be
cancelled by the operator $(\bar{\psi}\psi) F^3$. As a result, the
corresponding Wilson coefficient $\Sigma^{(3;-)}$ will contain some
logarithms.

In order to facilitate the presentation of the result let us introduce
the shorthand notations
\bea
l_{ij\ldots} &\equiv& l_i + l_j + \ldots \\
s_{kk'}(l_i,l_j) &\equiv& 
k\cdot l_i \ k'\cdot l_j - k\cdot l_j \ k'\cdot l_i
\\
s_{kk'}(l_i,l_j,l_k) &\equiv& 
k\cdot l_i \ k'\cdot l_j + k\cdot l_j \ k'\cdot l_j 
+ k\cdot l_j \ k'\cdot l_k 
\eea
and the  auxiliary function $f_{\rm aux}$ 
\bea
\lefteqn{f_{\rm aux}(k,k';l_1,l_2,l_3) \equiv
\frac{1}{4\ k\cdot l_1\ k\cdot l_{12}\ k\cdot l_{123}\
  k'\cdot l_1\ k'\cdot l_2\  k'\cdot l_3}} && \label{fauxdef} \\ 
&+& \frac{(k\cdot l_2)^2}{2\ k\cdot l_1\ k\cdot l_{12}\ k\cdot l_{123}\
k'\cdot l_2\ s_{kk'}(l_1,l_2)\ s^2_{kk'}(l_1,l_2,l_3)} 
\nonumber \\
&-& \frac{k\cdot l_2}{k\cdot l_{123}} \left(
\frac{1}{s_{kk'}(l_1,l_2)\ s^2_{kk'}(l_1,l_2,l_3)} +
\frac{1}{s^2_{kk'}(l_1,l_2)\ s_{kk'}(l_1,l_2,l_3)} \right)
\log\left(\frac{k\cdot l_1}{k\cdot l_{12}}\right)
\nonumber
\eea
The explicit form of the matching coefficient can then be written as
\bea
  \lefteqn{\Sigma^{(3;-)}_{\mu_1\nu_1\mu_2\nu_2\mu_3\nu_3}(k,k'|l_1,l_2,l_3)
  = \frac{1}{2^3}\ \big(k_{\mu_1}k'_{\nu_1}-k'_{\mu_1}k_{\nu_1}\big)
 \big(k_{\mu_2}k'_{\nu_2}-k'_{\mu_2}k_{\nu_2}\big)
 \big(k_{\mu_3}k'_{\nu_3}-k'_{\mu_3}k_{\nu_3}\big) }
  && \label{match_lin} \\ 
&& \times \
\bigg( f_{\rm aux}(k,k';l_1,l_2,l_3) - f_{\rm aux}(k,k';l_1,l_3,l_2)
-  f_{\rm aux}(k',k;l_1,l_2,l_3) + f_{\rm aux}(k',k;l_1,l_3,l_2)
\bigg) \nonumber 
\eea
We should mention that the form of the matching coefficient is not
unique, since it only enters in a symmetric combination in any
calculation. The important point is that the explicit form of the
matching coefficient \Eqn{match_lin} is compatible with the estimate
given in \Eqn{wilson_estimate}.

Let us now consider a second path. In order to evaluate the expression
given in \Eqn{sig2} we choose the following path. We start at $z_1$ and
follow a linear path to the intermediate point $w^{\alpha_3}=
z_1^{\alpha_3}+l_3^{\alpha_3}\ l_3\cdot(z_2-z_1)/l_3\cdot l_3$ 
from where we follow again a linear path to $z_2$. This path has the
advantage that
\be
\int_{z_1}^{z_2}d\omega^{\alpha_3}\ e^{-il_3\omega} =
i\, \frac{l_3^{\alpha_3}}{l_3\cdot l_3} 
\left( e^{-il_3z_2} - e^{-il_3z_1}  \right) +
 z_2^{\alpha_3}-z_1^{\alpha_3} - 
l_3^{\alpha_3}\ \frac{l_3\cdot(z_2-z_1)}{l_3\cdot l_3} 
\label{split_path}
\ee
and, therefore, \Eqn{sig2} will not yield logarithms upon integration.
However, the presence of terms proportional to $l_3^{\alpha_3}$ in
\Eqn{split_path} results in a more complicated tensor structure of the
Wilson coefficient. Instead of being proportional to $k_{\mu_i}$ or
$k'_{\nu_i}$ only, the matching coefficient will also contain tensor
structures with one of the fermion momenta replaced by a gluon
momentum. The explicit result is
\bea
  \lefteqn{\Sigma^{(3;-)}_{\mu_1\nu_1\mu_2\nu_2\mu_3\nu_3}
  (k,k'|l_1,l_2,l_3)  =  }
  && \label{match_split} \\
&&
\frac{1}{2^3}\ \big(k_{\mu_1}k'_{\nu_1}-k'_{\mu_1}k_{\nu_1}\big)
 \big(k_{\mu_2}k'_{\nu_2}-k'_{\mu_2}k_{\nu_2}\big)
 \big(k_{\mu_3}k'_{\nu_3}-k'_{\mu_3}k_{\nu_3}\big) 
 \Sigma^{(3;L^-)}(k,k'|l_1,l_2,l_3) 
\nonumber \\
&+&
 \frac{1}{2^3}\ \sum_{\sigma}\ (-1)^{\sigma}\
 l_{1\, \mu_1} k'_{\nu_1}
 \big(k_{\mu_2}k'_{\nu_2}-k'_{\mu_2}k_{\nu_2}\big)
 \big(k_{\mu_3}k'_{\nu_3}-k'_{\mu_3}k_{\nu_3}\big) 
\Sigma^{(3;SL^-)}(k,k'|l_1,l_2,l_3)
\nonumber
\eea
where the sum consists of twelve terms that are obtained by taking all
permutations of $\{1,2,3\}$ as well as the interchange terms
$k\leftrightarrow k'$. The factor $(-1)^{\sigma}$ denotes the sign of
the permutation of $\{1,2,3\}$. The explicit form of the scalar
coefficients is
\bea
\lefteqn{\Sigma^{(3;L^-)}(k,k'|l_1,l_2,l_3) = } &&
\\ && 
\frac{1}{2\, k\cdot l_2\  k\cdot l_3\  k'\cdot l_2\  k'\cdot l_3\ } 
\Bigg( \frac{1}{k\cdot l_1\ k\cdot l_{12}} 
 - \frac{1}{k'\cdot l_{12}\ k\cdot l_1 } \Bigg)
\nonumber \\
\nonumber \\
\lefteqn{\Sigma^{(3;SL^-)}(k,k'|l_1,l_2,l_3) = } &&
\\ && 
\frac{1}{3\ l_1\cdot l_1\ k\cdot l_3\ k'\cdot l_3} 
\Bigg( \frac{2}{k\cdot l_2\ k'\cdot l_2\ k'\cdot l_3 }
-  \frac{1}{k\cdot l_{12}\  k'\cdot l_1\ k'\cdot l_{12}} 
\Bigg)
\nonumber
\eea
Note that all terms in $\Sigma^{(3;-)}$ that are proportional to one of
the gluon momenta are path dependent and will be cancelled by
corresponding terms coming from the $(\bar{\psi}\psi) F^2$
operator. Therefore, as for the linear path, the explicit form of the
matching coefficient is in agreement with \Eqn{wilson_estimate}.



\end{document}